\documentclass[12pt]{article}
\usepackage{nicefrac}
\usepackage{amsfonts}
\usepackage{amsmath}
\usepackage{amsthm}

\advance \topmargin by -\headheight
\advance \topmargin by -\headsep
     
\evensidemargin \oddsidemargin
\def\rf#1{(\ref{eq:#1})}
\def\lab#1{\label{eq:#1}}
\def\nonu{\nonumber}
\def\br{\begin{eqnarray}}
\def\er{\end{eqnarray}}
\def\be{\begin{equation}}
\def\ee{\end{equation}}

\def\ba{\be\begin{array}{c}}
\def\ea{\end{array}\ee}

\def\lb{\lbrack}
\def\rb{\rbrack}

\def\({\left(}
\def\){\right)}
\def\v{\vert}                     
\def\lskip{\vskip\baselineskip\vskip-\parskip\noindent}

\def\bc{\begin{center}}
\def\ec{\end{center}}

\newcommand{\dder}[2]{\frac{d{#1}}{d{#2}}}
\newcommand{\pder}[2]{\frac{\partial{#1}}{\partial{#2}}}
\def\a{\alpha}
\def\b{\beta}

\def\d{\delta}

\def\eps{\epsilon}

\def\g{\gamma}

\newcommand{\h}{\frac{1}{2}}
\def\l{\lambda}

\def\p{\phi}

\def\pa{\partial}
\def\pr{\prime}

\def\cD{{\cal D}}

\def\cG{{\cal G}}

\def\cK{{\cal K}}
\def\cL{{\cal L}}
\def\cM{{\cal M}}

\newcommand{\ct}[1]{\cite{#1}}
\newcommand{\bi}[1]{\bibitem{#1}}
\def\cgh{{\widehat {\cal G}}}
\numberwithin{equation}{section}

\begin{document}


\begin{center}
{\large\bf  $\mathbf{N=2}$ and $\mathbf{N=4}$ Supersymmetric }
\end{center}
\begin{center}
{\large\bf mKdV and sinh-Gordon Hierarchies   }
\end{center}
\normalsize
\vskip .4in

\begin{center}
 H. Aratyn

\par \vskip .1in \noindent
Department of Physics \\
University of Illinois at Chicago\\
845 W. Taylor St.\\
Chicago, Illinois 60607-7059\\
\par \vskip .3in

\end{center}

\begin{center}
J.F. Gomes, L.H. Ymai
and A.H. Zimerman

\par \vskip .1in \noindent
Instituto de F\'{\i}sica Te\'{o}rica-UNESP\\
Rua Pamplona 145\\
01405-900 S\~{a}o Paulo, Brazil
\par \vskip .3in

\end{center}

\begin{abstract}
Integrable models with higher $N=2$ and $N=4$
supersymmetries are formulated on reductions of  
twisted loop superalgebras  $\widehat{ sl}(2|2)$ and  
$\widehat{sl}(4|4) $
endowed with principal gradation.
In case of the 
$\widehat{ sl}(4|4)$ loop algebra 
a sequence of progressing reductions
leads both to the $N=4$ and $N=2$ supersymmetric mKdV and 
sinh-Gordon equations.

The reduction scheme is induced by twisted 
automorphism and allows  via dressing approach 
to associate to each symmetry flow of half-integer
degree a supersymmetry transformation involving only local 
expressions in terms of the underlying fields.

\end{abstract}

\section{\sf Introduction}
We  further develop the systematic construction of supersymmetric integrable hierarchies 
in terms of the algebraic formalism proposed  in reference \ct{npskdv}.
In particular, we present a method leading to 
integrable models invariant under $N=2$ and $N=4$ extended 
supersymmetry realized in a local way (only expressions local in the
underlying fields are involved). The method consists
of the following key steps.
The basic Lie algebraic objects in the scheme 
are the loop algebras $\widehat{ sl}(2|2)$ and 
$\widehat{ sl}(4|4)$  endowed with a principal gradation
and a semisimple element $E$ of degree one.
The first step defines a symmetry structure of integrable
models under construction.  It involves 
dressing construction to associate
symmetry transformations of $\widehat{ sl}(2|2)$ and 
$\widehat{ sl}(4|4)$ loop algebras
to elements of the centralizer $\cK$ of $E$.
In this scheme the isospectral deformations defining
hierarchies of nonlinear flow equations correspond to 
the center of this centralizer. 
The rest of the centralizer
becomes associated to (in general non-commuting) additional
symmetry algebra of the underlying model.
Especially, in models with principal gradation, which we are considering,
the symmetry transformations corresponding to elements of 
the centralizer with the half-integer grades give raise 
to  supersymmetry transformations.  
The identity element belongs to the $\widehat{ sl}(2|2)$ and 
$\widehat{ sl}(4|4)$ loop algebras
and that has a consequence of doubling number of the 
supersymmetry generators
since squares of supersymmetry generators can be considered 
equivalent as long as they differ by identity element only.
In general, graded loop algebras with principal gradation
lead to non-local expressions for the supersymmetry 
transformations (e.g. the $N=2$ super mKdV
derived in \ct{npskdv} from the $\widehat{sl}(2|1)$ algebra).  
The non-locality can be traced back 
to a presence of a kernel 
term in the Lax operator required for
consistency in general case.
It is shown here that such term can be removed by a very 
judicious choice
of an underlying  relevant subalgebra within the loop structure. 
The key property of our reduction scheme 
is that it leads to reduced loop subalgebra
with image of $E$ containing only 
Cartan generators among even grade elements.
The emerging abelian structure ensures that the supersymmetry 
transformations remain local thus allowing us to circumvent
problem with non-local realization of supersymmetry.

We employ  this framework to construct equations of
motion for the generalized  mKdV hierarchies invariant under 
supersymmetry without non-local terms.
In particular, we derive the  equations of
motion for the $N=2$ supersymmetric mKdV and  sinh-Gordon models  
and show that they correspond to positive and negative grade time evolutions 
respectively and therefore  to the same hierarchy.

Construction of $N=4$ models involves two  subsequent 
reductions. The first one, reduces the algebra 
$\widehat{ sl}(4|4)$ into a semidirect product of $\widehat{U}(1)$ 
and $\widehat{sl}(2|2)\otimes \widehat{sl}(2|2)$. 
In the second reduction the relevant subalgebra is obtained
by  independently reducing each $\widehat{sl}(2|2)$ within its loop structure.

We were inspired by a number of papers devoted to
a supersymmetric generalization of the Drinfeld-Sokolov
construction. Inami and Kanno~\ct{Inami:1990ic,Inami:1991xj}
were first to
consider a class of affine superalgebras with fermionic simple roots
with the principal gradation.
Delduc and Gallot~\ct{Delduc-Gallot} and soon after 
Madsen and Miramontes \ct{Madsen:1999ta}
realized importance of fermionic elements in the kernel which square 
to $E$ for construction of a supersymmetric integrable hierarchy of 
the Drinfeld-Sokolov type.

Another set of related developments involves presence of 
non-local charges.
Kersten \ct{Kersten:1988ic} and Dargis
and Mathieu \ct{Dargis:1993jf,Mathieu:talk} obtained an infinite sequence
of non-local (bosonic and fermionic)
flow equations and conserved quantities in the framework of
supersymmetric KdV equation (see also \ct{DasPop}). 
These flows emerge in algebraic
formalism from (fermionic or bosonic) elements in $\cal K$ with 
non-negative grade as it was also observed in \ct{Madsen:1999ta}
and later in \ct{npskdv}.

There exists an extensive literature (see e.g. \ct{Popowicz:1995jf,
Sorin:2002mq,Delduc:2002nd,Nissimov:2001cq,Ivanov:1999ab,Delduc:1999br,
Delduc:1996mx,Ivanov:1996av})
devoted to construction of integrable models with 
extended supersymmetry based on a superspace. Such attempts
use as a rule covariant derivatives in expressions for the
Lax operator to ensure the supersymmetry invariance.
In our attempt supersymmetry emerges from structure of
symmetry transformations induced by fermionic directions
in the underlying superalgebra.

The paper is organized as follows. In Section \ref{section:dressing},
the main features of the dressing approach are described in a way which sets a
scene for supersymmetry invariant formulation of integrable hierarchies
of evolution flows.
Section \ref{section:reduce22} obtains $N=2$ mKdV and sinh-Gordon 
equations as zero-curvature conditions based on a subalgebra of 
$\widehat{ sl} (2 \v 2)$.
Extension of the above reduction process is applied in Section
\ref{section:n4mkdv} to the $\widehat{ sl}(4|4)$ algebra
resulting in $N=4$ mKdV and sinh-Gordon equations.
The $N=2$ mKdV and sinh-Gordon 
equations  of Section \ref{section:reduce22} can be reproduced 
by setting some of the fields of $N=4$ models to zero.
We offer some concluding remarks in Section \ref{section:conlude}.
Technical details of $\widehat{ sl}(4|4)$ algebra are relocated
to Appendix \ref{section:sl44algebra}.

\section{\sf Dressing Formalism}
\label{section:dressing}
In this Section we give a brief account of
the dressing approach to the integrable models.
 
We will be working with a loop algebra  $\cgh$ 
endowed with the principal gradation defined by a grading operator 
$Q$ to be given below for each specific model.
The gradation induces decomposition into graded
subspaces $ \cgh= \oplus_{n \in {\mathbb Z}} \, \cgh_n$ with  $\cgh_n$
such that $\lb Q \, , \, \cgh_n \rb = n\, \cgh_n$.

Another fundamental object in this setting
is a semisimple element $E$
of grade one.
The  Kernel $\cK$ of $E $ 
is a subalgebra 
of all elements commuting with $E$.
Elements of $\cK$  generate algebra of symmetries
commuting with isospectral deformations 
\ct{Aratyn:2000sm}. In the constructed models the
supersymmetry transformations belong to $\cK$ and carry
the half-integer grading.

Recall, that the 
center of kernel $\cK$ is defined 
as $C(\cK) = \{ x \in \cK\, \big\v \, [x,y]=0 \;\, \forall y \in \cK \}$.
Elements of $C(\cK)$
of grade $n \in {\mathbb Z}$  are denoted as 
$E^{(n)}$ and in this notation $E \equiv E^{(1)}$.
$E$ induces 
decomposition ${\hat \cG}= \cK \oplus \cM$
where $\cM$  is an 
image of the adjoint operation 
$ {\rm ad} (E) X = \left[ E, X \right]$.

To every positive grade
element $K_i  $ in $\cK$  one  associates
a transformation $\d_{K_i}$ through a  map  :
\be
K_i \in \cK \; \longrightarrow \;
\d_{K_i}\Theta = \(\Theta K_i \Theta^{-1}\)_{-} \Theta \, ,
\lab{kirheq}
\ee
where $(X)_{-}$ is a projection of $X \in \cgh$ on
a strictly negative  part of $X$ in 
$\cgh_{<} = \oplus_{n=-1}^{-\infty} \, \cgh_n$
and 
$\Theta$, the dressing matrix, is 
an exponential in $\cG_{<}$ :
\be
\Theta = \exp \( \sum_{i<0} \theta^{(i)} \)
= \exp \( \theta^{(-1)} +
\theta^{(-{2})} +{\ldots} \) \, .
\lab{dressmat}
\ee
The map $K_i \, \to \, \d_{K_i}$ is a homomorphism \ct{Aratyn:2000sm}:
\be
\left[ \d_{K_i} \,, \,  \d_{K_j} \right] \Theta =
\d_{\left[ {K_i} ,  {K_j} \right]} \Theta \, .
\lab{dhomo}
\ee
It allows us to identify $\cK$ with algebra of symmetry transformations.

For $E^{(n)} \in C(\cK)$,  
the corresponding flows
$\d_{E^{(n)}} $ define through the map \rf{kirheq}
the isospectral deformations of the model. 
They obviously commute among themselves 
and we  denote them as partial derivatives $ \pa /\pa t_n$.
By definition :
\be
\pder{}{t_n} \Theta (t)= 
\(\Theta E^{(n)} \Theta^{-1}\)_{-} \Theta (t)\, .
\lab{rheq}
\ee
The Lax operator $L$ is obtained by 
dressing the isospectral flow with $n=1$ in equation \rf{rheq}.
Let us identify $t_1$ with the space variable $x$.
Then :
\be
\begin{split}
\pa_x  (\Theta) &=\(\Theta E \Theta^{-1}\)_{-} \Theta
= \big\lbrack \Theta E \Theta^{-1} - \(\Theta E \Theta^{-1}\)_{+}
\big\rbrack \Theta  \\
&= \Theta  E - \( E+ \left[ \theta^{(-1)} , E \right] \) \Theta\\
&= \Theta  E - \( E+ A_0 \) \Theta
\end{split}
\lab{rht1sl2}
\ee
where $(\cdot{})_{+}$ denotes projection on a positive
subalgebra $\cgh_{\geq } =\oplus_{n=0}^{\infty} \, \cgh_n$. 
Note that $ A_0 = \left[ \theta^{(-1)} , E \right] $
is clearly in $\cM$ and of grade zero.
This leads to the dressing expression for the Lax operator 
$L$ :
\be
\Theta\( \pa_x+ E \)\Theta^{-1} =   \pa_x+ E+ A_0  
= L \, .
\lab{dreslaxsl2}
\ee
Similarly, for 
higher flows we obtain
\be
 \Theta  \( \pder{}{t_n}+ E^{(n)}\) \Theta^{-1}
= \pder{}{t_n}+ E^{(n)}+ \sum_{i=0}^{n-1} D^{(i)}_n  
\lab{dresbn}
\ee
with coefficients $D^{(i)}_n $ which can be found from identities:
\[ \(\Theta E^{(n)} \Theta^{-1}\)_{+} =
E^{(n)}+ \sum_{i=0}^{n-1} D^{(i)}_n \, .
\]
These dressing relations give rise to the zero-curvature conditions
\be
\left[ \pa_x+ E + A_0\, ,\, \pder{}{t_n}+ E^{(n)}+ \sum_{i=0}^{n-1} D_n^{(i)}
\right]
= \Theta \left[ \pa_x+ E\, ,\, \pder{}{t_n}+ E^{(n)} \right]
\Theta^{-1 }  =0 \, .
\lab{z-curva}
\ee
The structure of the Lax operator
changes when terms with the half-integer grades 
appear in $\cgh = \oplus_{n \in {\mathbb Z}} \, \cgh_{n/2}$.
With these terms being present in the exponent of the 
dressing matrix equation \rf{dressmat} generalizes to : 
\be 
\Theta = \exp \( \sum_{i<0} \theta^{(i)} \)
= \exp \( \theta^{(-1/2)} +  \theta^{(-1)} +
\theta^{(-\nicefrac{3}{2})} +{\ldots} \) 
\lab{dressmathalf}
\ee
and the form of the Lax operator obtained by 
the dressing procedure changes as follows :
\be
\begin{split}
\pder{}{t_1} (\Theta) &=\(\Theta E \Theta^{-1}\)_{-} \Theta
= \big\lbrack \Theta E \Theta^{-1} - \(\Theta E \Theta^{-1}\)_{+}
\big\rbrack \Theta  \\
&= \Theta  E + \( E+ \left[ \theta^{(-1)} , E \right] +
\left[ \theta^{(-1/2)} , E \right] + \h \left[ \theta^{(-1/2)} , \left[
\theta^{(-1/2)} , E \right]\right] \) \Theta \\
&= \Theta  E + \( E+ A_0 + A_{1/2} +k_0 \) \Theta \, .
\end{split}
\lab{rht1}
\ee
Here 
\br
A_0 &=& \left[ \theta^{(-1)} , E \right] + 
\h \left[ \theta^{(-1/2)} , \left[
\theta^{(-1/2)} , E \right]\right] \Big\v_{\cM}  \, \in \,\cM \lab{azero}\\
A_{1/2}&=& \left[ \theta^{(-1/2)} , E \right] \, \in \,\cM
\lab{aonehalf}\\
k_0 &=& \h \left[ \theta^{(-1/2)} , \left[
\theta^{(-1/2)} , E \right]\right] \Big\v_{\cK}
\, \in \,\cK
\lab{kzero}
\er
where $ \Big\v_{\cK}$ and $\Big\v_{\cM}$
denote projections on the kernel ${\cK}$ and image ${\cM}$, 
respectively.
This shows that, in case of a half-integer grading, 
a general expression for the Lax 
operator is
\be 
\cL = \pa_x + E+ A_0 + A_{\nicefrac{1}{2}} +k_0 \,.
\lab{laxoper}
\ee
The unconventional grade zero term  $k_0$ residing
in $\cK$ appears here 
due to the  half-integer grading (encountered
in case of $\widehat{ sl} (n|m)$ algebras with principal gradation).

The presence of $k_0$ in the Lax operator $L$ presents a problem for our
formalism as it causes 
non-locality in the supersymmetry transformations 
\ct{npskdv} as follows from the zero-curvature equations
we will describe below.
A remedy employed in this paper is to work with subalgebras 
obtained by 
reductions in which the term 
$\left[ \theta^{(-1/2)} , \left[
\theta^{(-1/2)} , E \right]\right] $ vanishes automatically.
This paper describes such construction for $\widehat{ sl}(n \v n)$ 
with $n=2,4$.

Consider now the case of a kernel $\cK$ which contains 
a constant grade one-half element $D^{(\nicefrac{1}{2})}$.
According to \rf{kirheq} this term gives rise to the symmetry 
flow
\be
\pa_{\nicefrac{1}{2}} \Theta \equiv \d_{D^{(\nicefrac{1}{2})}}\Theta 
= \( \Theta D^{(\nicefrac{1}{2})} 
\Theta^{-1} \)_{-}
 \Theta \,.
\lab{defsusy}
\ee
As shown in \ct{npskdv}, $\pa_{\nicefrac{1}{2}}$-flow enters the zero-curvature
equation :
\be
 \left[ \pa_x +E +A_0 +A_{\nicefrac{1}{2}} \, , \, \pa_{\nicefrac{1}{2}} + 
D^{(0)}+D^{(\nicefrac{1}{2})} \right]=0\, ,
\lab{zchalf}
\ee
where
\[ D^{(0)} =  \left[ \theta^{(-\nicefrac{1}{2})}, 
D^{(\nicefrac{1}{2})}  \right] \, .
\]
In presence of the half-integer grading
the zero-curvature condition \rf{z-curva} gives way to
(for reductions with $k_0=0$) :
\be
 \left[ \pa_x +E +A_0 +A_{\nicefrac{1}{2}}\, ,\, \pder{}{t_n} + 
E^{(n)}+\sum_{k=1}^n \(D_n^{(k)} +D_n^{(k-\nicefrac{1}{2})}\) 
\right]=0
\lab{zc1}
\ee
As described in \ct{npskdv} this zero-curvature approach extends to 
the negative flows $\pa/\pa t_{-n}$. 
The following three Lax operators :
\br
\cD_{+1}&=& \cL= 
\pa_x + A_0 +A_{\nicefrac{1}{2}}+E
\lab{cdplus}\\
\cD_{+\nicefrac{1}{2}}&=& \pa_{\nicefrac{1}{2}}  + D^{(0)} +
D^{(\nicefrac{1}{2})}
\lab{cdhalf}\\
\cD_{-1}&=& \pa_{-1} + B \jmath_{-\nicefrac{1}{2}} B^{-1} +B E^{(-1)} B^{-1}
\lab{cdminus}
\er
appear prominently in the dressing analysis.
Here $B$ is a non-singular matrix 
of grade zero and $ \jmath_{-\nicefrac{1}{2}}$ is an element 
in $\cM$ of grade $-1/2$.
The flows $\pa_{\nicefrac{1}{2}}, \pa_{-1} = \pa /\pa t_{-1}$
and $\pa_x=\pa / \pa t_1$
are all commuting among themselves, being associated to
$D^{(\nicefrac{1}{2})}$, $E^{(-1)}$ and $E$, respectively.
By standard dressing argument this commutativity ensures
the zero-curvature equations :
\br
\left[ \cD_{+\nicefrac{1}{2}} \, , \, \cD_{-1} \right]&=&0
\lab{cdhalfcdminus}\\
\left[ \cD_{+\nicefrac{1}{2}} \, , \, \cD_{+1} \right]&=&0
\lab{cdhalfcdplus}\\
\left[ \cD_{+1} \, , \, \cD_{-1} \right]&=&0 \, .
\lab{cdpluscdminus}
\er
The brackets with $\cD_{+\nicefrac{1}{2}}$ act as compatibility equations which
define supersymmetry transformations and ensure invariance of
equations of motion (\rf{cdpluscdminus}), i.e.
\be
 \left[ \pa_x +E +A_0+A_{\nicefrac{1}{2}}\, ,\,  \pa_{-1} + 
B \jmath_{-\nicefrac{1}{2}} B^{-1} +B E^{(-1)} B^{-1}
  \right]=0
\lab{zcneg}
\ee
under the supersymmetry transformation.
The  $-1$ grade component of the zero-curvature relation \rf{zcneg}
\be
\pa_x \(B E^{(-1)} B^{-1}\) + 
 \left[ A_0 ,\, B E^{(-1)} B^{-1}  \right]=0
\lab{grm1comp}
\ee
is automatically satisfied for 
\be
A_0 = -\pa_{x} B\, B^{-1}\, .
\lab{idenaz}
\ee
Equation 
\rf{cdhalfcdminus} is explicitly given by
\[
\left[ \pa_{\nicefrac{1}{2}}  + D^{(0)} +D^{(\nicefrac{1}{2})}, \pa_{-1} + 
B \jmath_{-\nicefrac{1}{2}} B^{-1} +B E^{(-1)} B^{-1}
  \right]=0 \, .
\]
Its $-1$  grade component is equal to 
\[
\pa_{\nicefrac{1}{2}}\(B E^{(-1)} B^{-1}\)  +\left[ D^{(0)} ,
\, B E^{(-1)} B^{-1} \right] =0 \,.
\]
which has the obvious solution :
\be
D^{(0)}=  -\pa_{\nicefrac{1}{2}} B\, B^{-1} \, .
\lab{d0sol}
\ee
With solutions \rf{idenaz} and \rf{d0sol} we can rewrite $\cD_{+1}$ and
$\cD_{+\nicefrac{1}{2}} $ as
\br
\cD_{+1}&=& \cL= 
\pa_x -\pa_x B \,B^{-1}+ A_{\nicefrac{1}{2}} +E
\lab{cdplusa}\\
\cD_{+\nicefrac{1}{2}}&=& \pa_{\nicefrac{1}{2}}  -\pa_{\nicefrac{1}{2}} B\,
B^{-1} +D^{(\nicefrac{1}{2})} \, .
\lab{cdhalfa}
\er
We now write explicitly all the zero-curvature equations in 
components.
Equation \rf{cdhalfcdminus} implies
\br
\pa_{\nicefrac{1}{2}} \jmath_{-\nicefrac{1}{2}} &=& \left[ E^{(-1)} \, , \,  B^{-1} D^{(\nicefrac{1}{2})} B \right]
\lab{pahjmh}\\
\pa_{-1} \( \pa_{\nicefrac{1}{2}} B\, B^{-1} \)  &=& \left[
 B \jmath_{-\nicefrac{1}{2}} B^{-1} \, , \,  D^{(\nicefrac{1}{2})} \right] \, .
\lab{pampahbb}
\er
{}From \rf{cdhalfcdplus} we derive 
\br
\left[ E \, , \,  \pa_{\nicefrac{1}{2}} B\, B^{-1} \right] 
&=& 
\left[ A_{\nicefrac{1}{2}} \, , \,  D^{(\nicefrac{1}{2})} \right] 
 \lab{jhalfdhalf}\\
\pa_{\nicefrac{1}{2}}\, A_{\nicefrac{1}{2}} &=&  \left[\pa_{\nicefrac{1}{2}} B\, B^{-1}\, , \,  A_{\nicefrac{1}{2}} \right]
- \left[\pa_{x} B\, B^{-1}\, , \, D^{(\nicefrac{1}{2})} \right]\, .
\lab{pahjph}
\er
Finally, equation \rf{cdpluscdminus} yields
\br
\pa_{x}\, \jmath_{-\nicefrac{1}{2}} 
&=&  \left[E^{(-1)}\, , \,   B^{-1}  A_{\nicefrac{1}{2}}  B \right]
\lab{paxjmh}\\
\pa_{-1}\, A_{\nicefrac{1}{2}} &=&  \left[E\, , \,   B \jmath_{-\nicefrac{1}{2}}  B^{-1}  \right]
\lab{pamjph}\\
\pa_{-1} \( \pa_{x} B\, B^{-1} \)  &=& 
\left[B E^{(-1)} B^{-1}\, , \,E \right]
\, + \,\left[ B \jmath_{-\nicefrac{1}{2}} B^{-1}  \, , \,
A_{+\nicefrac{1}{2}} \right] \,.
\lab{pampaxbb}
\er
The zero curvature condition
\[
 \left[ \cD_{\nicefrac{1}{2}} \, , \, \cD_n \right]=0
\]
for the higher flows generators
\[
\cD_n = \pder{}{t_n} + E^{(n)}+
 \sum_{k=1}^n \(D_n^{(k-1)} +D_n^{(k-\nicefrac{1}{2})}\) \, , 
\]
ensures invariance of higher flows 
under the supersymmetry transformation.

\section{\sf Reduction of $\widehat{ sl} (2\v 2)$ Algebra and $N=2$
mKdV Supersymmetric System}
\label{section:reduce22}
Let us define basic objects.
The principal grading for the $\widehat{ sl} (2|2)$ algebra
is defined in terms of the operator
\be
Q= \l \dder{}{\l} + \h (\a_1+\a_3)\cdot H
= \l \dder{}{\l}+
\h \left[ 
{\begin{array}{cc|cc} 
1&0&0&0 \\
0&-1&0&0\tabularnewline\hline
0&0&-1&0\\
0&0&0&1
\end{array}}
 \right]  \, .
\lab{gradop}
\ee
Grade one semisimple element $E$ is chosen as
\be
E = (E_{\a_1}^{(0)} + E_{-\a_1}^{(2)}) + 
(E_{\a_3}^{(2)} + E_{-\a_3}^{(0)}) + I^{(1)} 
=\left[ 
{\begin{array}{cc|cc} 
\l&1&0&0 \\
\l^2&\l&0&0\tabularnewline\hline
0&0&\l&\l^2\\
0&0&1&\l
\end{array}}
 \right] \, .
\lab{esl22}
\ee
The odd (fermionic) part of the kernel of $E$ consists of
\be
\cK_f= \{ f_i^{(n+\frac{1}{2})} , i=1,{\ldots} ,4, \;\;\;
n \in {\mathbb Z} \}
\lab{ckf22}
\ee
with
\br
f_1^{(n+\h)} &=&  (E_{\a_1+\a_2}^{(n-\h)} + E_{-\a_1-\a_2}^{(n+\frac{3}{2})}) + 
(E_{\a_2+\a_3}^{(n+\frac{3}{2})} + E_{-\a_2-\a_3}^{(n-\h)}), \nonu \\
f_2^{(n+\h)} &=&  (-E_{\a_1+\a_2}^{(n-\h)} + E_{-\a_1-\a_2}^{(n+\frac{3}{2})}) + 
(-E_{\a_2+\a_3}^{(n+\frac{3}{2})} + E_{-\a_2-\a_3}^{(n-\h)}), \nonu \\
f_3^{(n+\h)} &=&  (E_{\a_1+\a_2+\a_3}^{(n+\h)} + E_{-\a_1-\a_2-\a_3}^{(n+\h)}) + 
(E_{\a_2}^{(n+\h)} + E_{-\a_2}^{(n+\h)}), 
\nonu \\
f_4^{(n+\h)} &=&  (-E_{\a_1+\a_2+\a_3}^{(n+\h)} + E_{-\a_1-\a_2-\a_3}^{(n+\h)}) + 
(-E_{\a_2}^{(n+\h)} + E_{-\a_2}^{(n+\h)}), 
\nonu \\
\er
while the bosonic part 
$ \cK_b= \{ K_i^{(n)} , i=1,{\ldots} ,3,\;\;\;
n \in {\mathbb Z} \} $ of the kernel $\cK$
contains
\[ K_1^{(n)} = E_{\a_1}^{(n-1)} + E_{-\a_1}^{(n+1)},\;\;\;
K_2^{(n)} = E_{\a_3}^{(n+1)} + E_{-\a_3}^{(n-1)},\;\;\;
K_3^{(n)}  = \l^n I \,.
\]
where $I$ is an identity matrix, traceless with respect to the
supertrace.

Note, that in terms of generators from $ \cK_b$ we can write
\[ E= K_1^{(1)}+ K_2^{(1)}+ K_3^{(1)}.\]
Because $K_3$ commutes with the rest of the algebra the symmetry transformation
$\d_{K_3}$ defined according to \rf{kirheq} vanishes.
Thus, the flows generated by, i.e.
\[{\widetilde E}^{(n)}= K_1^{(n)}+ K_2^{(n)}- K_3^{(n)}\]
will give rise via relation \rf{kirheq} to flows  identical to 
the isospectral deformations generated  by $E^{(n)}$.
As far as symmetry transformations go we can therefore
identify ${E}^{(n)}$ with ${\widetilde E}^{(n)}$.
As we will see below this observation is crucial for extending
supersymmetry by incorporating as supersymmetry flows 
the flows generated by elements of $\cK_f$ which square to
${E}$ as well as those generated by elements 
which square to ${\widetilde E}^{(1)}$.

We change now the basis in $\cK_f$ to :
\begin{alignat}{2}
F_1^{(n+\h)} &= \frac{1}{\sqrt{2}}\( f_1^{(n+\h)} +f_3^{(n+\h)} \), 
&\qquad  F_2^{(n+\h)} &= 
\frac{1}{\sqrt{2}}\( f_2^{(n+\h)} +f_4^{(n+\h)} \), \nonu \\
F_3^{(n+\h)} &= \frac{1}{\sqrt{2}}\( f_1^{(n+\h)} -f_3^{(n+\h)} \), 
&\qquad  F_4^{(n+\h)} &= \frac{1}{\sqrt{2}}\( f_2^{(n+\h)} -f_4^{(n+\h)} \)\,. \nonu 
\end{alignat}
The advantage of using this basis is that its elements
generate fermionic flows that square to 
isospectral deformations generated  by $E^{(n)}$
(or ${\widetilde E}^{(n)}$)
according to anti-commutation relations
of $F_i^{(n+\h)}, \, i=1,{\ldots} ,4$ :
\br
\{F_i^{(m+\h)}, F_i^{(n+\h)}\} &=& 2(-1)^{i+1} E^{(m+n+1)},
\quad i=1,2
\lab{fisqe} \\
\{F_i^{(m+\h)}, F_i^{(n+\h)}\} &=& 2 (-1)^{i} {\widetilde E}^{(m+n+1)} \, .
\quad i=3,4
\lab{fisqte}
\er
The remaining anti-commutation relations
of $F_i^{(n+\h)}, \, i=1,{\ldots} ,4$ vanish.
In this sense $F_i^{(\h)}, \, i=1,{\ldots} ,4$ generate the
supersymmetry flows.

The fermionic part of the image $\cM$ of $E$ consists of
\be
\cM_f= \{ g_i^{(n+\frac{1}{2})} , i=1,{\ldots} ,4, \;\;\;
n \in {\mathbb Z} \}
\lab{cmg22}
\ee
where
\br
g_1^{(n+\h)} &=&  (E_{\a_1+\a_2}^{(n-\h)} + E_{-\a_1-\a_2}^{(n+\frac{3}{2})}) - 
(E_{\a_2+\a_3}^{(n+\frac{3}{2})} + E_{-\a_2-\a_3}^{(n-\h)}), \nonu \\
g_2^{(n+\h)} &=&  (-E_{\a_1+\a_2}^{(n-\h)} + E_{-\a_1-\a_2}^{(n+\frac{3}{2})}) +
(E_{\a_2+\a_3}^{(n+\frac{3}{2})} - E_{-\a_2-\a_3}^{(n-\h)}), \nonu \\
g_3^{(n+\h)} &=&  (E_{\a_1+\a_2+\a_3}^{(n+\h)} + E_{-\a_1-\a_2-\a_3}^{(n+\h)}) -
 (E_{\a_2}^{(n+\h)} + E_{-\a_2}^{(n+\h)}), 
\nonu \\
g_4^{(n+\h)} &=&  (-E_{\a_1+\a_2+\a_3}^{(n+\h)} + E_{-\a_1-\a_2-\a_3}^{(n+\h)}) + 
(E_{\a_2}^{(n+\h)} - E_{-\a_2}^{(n+\h)}), 
\nonu \\
\er
Again we will rather work with a different basis of $\cM_f$ 
given by
\begin{alignat}{2}
G_1^{(n+\h)} &= \frac{1}{\sqrt{2}}\( g_1^{(n+\h)} +g_3^{(n+\h)} \), 
&\qquad G_2^{(n+\h)} &= \frac{1}{\sqrt{2}}\( g_2^{(n+\h)} +g_4^{(n+\h)} \), \nonu \\
G_3^{(n+\h)} &= \frac{1}{\sqrt{2}}\( g_1^{(n+\h)} -g_3^{(n+\h)} \), 
&\qquad G_4^{(n+\h)} &= \frac{1}{\sqrt{2}}\( g_2^{(n+\h)} -g_4^{(n+\h)} \) \, . \nonu \\
\end{alignat}
There are four bosonic generators 
\begin{alignat}{2}
M_1^{(n)} &= \a_1\cdot H^{(n)}, &\qquad  M_2^{(n)} &= -E_{\a_1}^{(n-1)} + E_{-\a_1}^{(n+1)}, \nonu \\
M_3^{(n)} &=\a_3\cdot H^{(n)}, &\qquad 
M_4^{(n)} &= -E_{\a_3}^{(n+1)} + E_{-\a_3}^{(n-1)}, \nonu 
\end{alignat}
of the image
$\cM$ of $E$. Note, that $M_1^{(n)}$ and $M_3^{(n)}$ are in 
the Cartan subalgebra. 
All four bosonic generators of $\cM$ are reproduced by the 
following anti-commutations relations between fermionic members of 
$\cK$ and $\cM$ :
\br
\{  G_1^{(m+\h)},  F_i^{(n+\h)}\} &=& 2(-1)^{i+1}
M_i^{(m+n+1)},\quad i=1,2,3,4\nonu \\
\{  G_4^{(m+\h)},  F_i^{(n+\h)}\} &=& -2M_{5-i}^{(m+n+1)}    ,\quad
i=1,2,3,4   \,.\nonu 
\er
Using relations
\begin{alignat}{2}
  \left[E^{(m)},  G_1^{(n+\h)}\right] 
  & = 2 G_2^{(m+n+\h)}, &\qquad \left[E^{(m)}, G_2^{(n+\h)}\right]
  & = 2 G_1^{(m+n+\h)}  \nonu\\
 \left[E^{(m)},  G_3^{(n+\h)}\right] & = -2 G_4^{(m+n+\h)}
, &\qquad \left[E^{(m)},  G_4^{(n+\h)}\right] & = 
 -2 G_3^{(m+n+\h)} \nonu 
\end{alignat}
and 
\begin{alignat}{2}
  \left[E^{(m)},  M_1^{(n)}\right] 
  & = 2 M_2^{(m+n)}, &\qquad \left[E^{(m)}, M_2^{(n)}\right]
  & = 2 M_1^{(m+n)}  \nonu\\
 \left[E^{(m)},  M_3^{(n)}\right] & =  2M_4^{(m+n)},
 &\qquad \left[E^{(m)},   M_4^{(n)}\right] & = 
 2 M_3^{(m+n)} \nonu 
\end{alignat}
one can obtain the remaining anti-commutation relations between
$G_2^{(m+\h)}, G_3^{(m+\h)}$ and $F_i^{(n+\h)}$.

\subsection{\sf Reduction of $\widehat{ sl} (2 \v 2)$
and N=2 mKdV and Sinh-Gordon equations }
The $\widehat{ sl} (2 \v 2)$ algebra splits in two disjoint sets
of generators :
\begin{gather}
M_1^{(2n)}, M_3^{(2n)},\quad M_2^{(2n+1)}, M_4^{(2n+1)},\quad
K_1^{(2n+1)}, 
K_2^{(2n+1)},I^{(2n+1)} \nonu\\
 G_1^{(2n+\h)}, G_3^{(2n+\h)},
 F_2^{(2n+\h)},  F_4^{(2n+\h)},\lab{redsl22}\\
 G_2^{(2n+\frac{3}{2})},  G_4^{(2n+\frac{3}{2})},\quad 
 F_1^{(2n+\frac{3}{2})}, \quad  F_3^{(2n+\frac{3}{2})}\nonu
\end{gather}
and 
\begin{gather}
M_1^{(2n+1)}, M_3^{(2n+1)},\quad M_2^{(2n)}, M_4^{(2n)},\quad
K_1^{(2n)}, 
K_2^{(2n)},I^{(2n)} \nonu\\
 G_1^{(2n+\frac{3}{2})}, G_3^{(2n+\frac{3}{2})},
 F_2^{(2n+\frac{3}{2})},  F_4^{(2n+\frac{3}{2})},\nonu\\
 G_2^{(2n+\frac{1}{2})},  G_4^{(2n+\frac{1}{2})},\quad 
 F_1^{(2n+\frac{1}{2})}, \quad  F_3^{(2n+\frac{1}{2})}\nonu
\end{gather}
where $n \in {\mathbb Z}$.

Our reduction process is now facilitated by the fact that 
the set \rf{redsl22} of $\widehat{ sl} (2 \v 2)$ generators 
constitutes a subalgebra, which we denote by 
$\widehat{ sl}_0 (2 \v 2)$.
The $\widehat{ sl}_0 (2 \v 2)$ subalgebra includes $E=K_1^{(1)}+K_2^{(1)}+I^{(1)}$
and the image of $E$ within $\widehat{ sl}_0 (2 \v 2)$ contains only 
the Cartan generators $M_1^{(2n)}$ and $M_3^{(2n)}$
among even grade elements.  Notice that within 
the $\widehat{ sl}_0 (2 \v 2)$ subalgebra  the term 
$k_0 $ from \rf{kzero} vanishes identically.

We now associate the Lax operator 
\be 
L = \pa_x + E+ A_0 + A_{\frac{1}{2}}  \,
\lab{laxoper1}
\ee
to $\widehat{ sl}_0 (2 \v 2)$ subalgebra 
by choosing \[
A_0= u_1 M_1^{(0)}+ u_3 M_3^{(0)}, \quad 
A_{\h} = {\bar \psi}_1 G_1^{(\h)} + {\bar \psi}_3 G_3^{(\h)}
\]
which are elements of $\widehat{sl}_0 (2 \v 2)$
with grade zero and $\nicefrac{1}{2}$, respectively.

To find $N=2$ mKdV equations we need to solve the zero-curvature 
equation \rf{zc1} for $n=3$. It  is explicitly given by
\be
 \left[ \pa_x +E +A_0 + A_{\frac{1}{2}}, \pa_3 + 
D_3^{(0)}+D_3^{(\frac{1}{2})} +D_3^{(1)} + D_3^{(\frac{3}{2})} +D_3^{(2)}+ 
D_3^{(\frac{5}{2})}+E^{(3)}
 \right]=0
\lab{zc3}
\ee
with  $E^{(3)}= K_1^{(3)}+K_2^{(3)}+I_2^{(3)}$.

We found the following solution for the $D$'s within 
the $\widehat{ sl}_0 (2 \v 2)$ subalgebra :
\begin{alignat}{2}
 D^{(5/2)}_M &= \l^2 A_{\h}, &\qquad 
D^{(5/2)}_K &=0 {}\nonu \\
D^{(2)}_M&= \l^2 A_{0} , &\qquad D^{(2)}_K &= 0 \nonu\\
D^{(3/2)}_M &= \h  \( \pa_x  \psi_1\, G_2^{(\frac{3}{2})} -
\pa_x  \psi_3\, G_4^{(\frac{3}{2})}\)
, &\qquad 
D^{(3/2)}_K &= d_1 F_1^{(\frac{3}{2})}+ 
d_3 F_3^{(\frac{3}{2})}
\nonu \\
D^{(1)}_M&=  c_2 M_2^{(1)}  + c_4 M_4^{(1)}  
, &\qquad 
D^{(1)}_K&= a_1 K_1^{(1)} +a_2 K_2^{(1)}+
a_0 I_2^{(1)}
\nonu\\
D^{(1/2)}_M&=  \b_1 G_1^{(\frac{1}{2})} +\b_3 G_3^{(\frac{1}{2})} 
, &\qquad 
D^{(1/2)}_K&= \gamma_2 F_2^{(\frac{1}{2})}+\gamma_4 F_4^{(\frac{1}{2})}
\nonu\\ 
D^{(0)}_M&=  \d_1 M_1^{(0)} +\d_3 M_3^{(0)} 
, &\qquad 
D^{(0)}_K&=0 \nonu
\end{alignat}
where
\begin{alignat}{2}
d_1&=- \h \(\psi_1 u_1 +\psi_3 u_3  \), &\qquad 
d_3&= \h \(\psi_1 u_3 +\psi_3 u_1  \)\nonu\\
c_2&= \h u_1^{\pr}-\psi_1\(\psi_3u_3\) , &\qquad 
c_4&= \h u_3^{\pr}+\psi_1\(\psi_3u_1\) \nonu\\
a_0&=\h\(-\psi_1 \pa_x \psi_1 - \psi_3 \pa_x \psi_3\), &\qquad 
{}&{}\nonu\\
a_1&=\h\(\psi_1 \pa_x \psi_1 - \psi_3 \pa_x \psi_3- u_1^2\) 
, &\qquad 
a_2&=\h\(-\psi_1 \pa_x \psi_1 + \psi_3 \pa_x \psi_3-u_3^2\) \nonu\\
\b_1&= \frac{1}{4} \pa_x^2 \psi_1- 
\frac{1}{2} \psi_1  \( u_1^2+ u_3^2\)
-\h \psi_3  \(u_1u_3 \) , &\qquad 
\b_3&= \frac{1}{4} \pa_x^2 \psi_3- 
\frac{1}{2} \psi_3  \( u_1^2+ u_3^2\)
-\h \psi_1  \(u_1u_3 \) \nonu \\
\gamma_2&= \frac{1}{4} ( \psi_1 u_1^{\pr} - \psi_1^{\pr}  u_1
-\psi_3 u_3^{\pr} + \psi_3^{\pr}  u_3 ), &\qquad 
\gamma_4&= \frac{1}{4} ( -\psi_1 u_3^{\pr} +\psi_1^{\pr}  u_3
+\psi_3 u_1^{\pr} - \psi_3^{\pr}  u_1 )\nonu
\end{alignat}
and
\br
4\d_1&=&  \pa_x^2(u_1)- 2 u_1^3 +{3}
u_1 \( \psi_1 \pa_x \psi_1 - \psi_3 \pa_x \psi_3\) +{3}
u_3 \( -\psi_1 \pa_x \psi_3 + \psi_3 \pa_x \psi_1\) 
\lab{delt1}\\
4\d_3&=&  \pa_x^2(u_3)- 2 u_3^3 -{3}
u_3 \( \psi_1 \pa_x \psi_1 - \psi_3 \pa_x \psi_3\) 
-{3}
u_1 \( -\psi_1 \pa_x \psi_3 + \psi_3 \pa_x \psi_1\) 
\lab{delt3}
\er
This leads to the following equations of motion :
\br
4 \pa_3  \psi_1 &=& \pa_x^3 \psi_1- 
\frac{3}{2} \psi_1 \pa_x \( u_1^2+ u_3^2\)
- 3  \pa_x (\psi_1)  \( u_1^2+ u_3^2\)
-3 \psi_3 \pa_x \(u_1u_3 \) 
\lab{xi1eqsmot}
\\
4 \pa_3  \psi_3 &=&
\pa_x^3 \psi_3- 
\frac{3}{2} \psi_3 \pa_x \( u_1^2+ u_3^2\)
- 3  \pa_x (\psi_3)  \( u_1^2+ u_3^2\)
-3 \psi_1 \pa_x \(u_1u_3 \) 
\lab{xi3eqsmot}
\er
for fermionic and  :
\be
\pa_3 u_i = \pa_x \d_i, \quad i=1,3
\lab{ueqsmot}
\ee
for bosonic modes. These are the $N=2$ supersymmetric mKdV equations.
They are invariant under the supersymmetry transformations 
\be \pa_{\h} u_1 = 2 
\pa_x  \(-\psi_1 \eps_2 +\psi_3 \eps_4  \), \;\;\;\;
\pa_{\h} u_3 = 2 \pa_x \(-\psi_1 \eps_4 +\psi_3 \eps_2 \) 
\lab{susn2u}
\ee
and
\be
\pa_{\h} \psi_1 = u_1 \eps_2 - u_3 \eps_4 , \quad
\pa_{\h} \psi_3 = u_1 \eps_4 -u_3 \eps_2
\lab{susn2psi}
\ee
derived from the 
zero-curvature equation \rf{zchalf}
with
\br
D^{(\frac{1}{2})} &= &\eps_2 F_2^{(1/2)} +\eps_4 F_4^{(1/2)}
 \lab{dshalf}\\
D^{(0)} &= &2 \(-\psi_1 \eps_2 +\psi_3 \eps_4  \) M_1^{(0)} 
+2 \(-\psi_1 \eps_4 +\psi_3 \eps_4 \)  M_3^{(0)} \,.
\lab{dzm}
\er
Applying the transformations \rf{susn2u}-\rf{susn2psi}
twice we obtain:
\begin{alignat}{2}
\pa_{\h}^2 u_1 &= -4 \pa_x u_3 \,\eps_2 \,\eps_4, &\qquad 
\pa_{\h}^2 u_3 &= -4 \pa_x u_1 \,\eps_2 \,\eps_4\, , \nonu\\
\pa_{\h}^2 \psi_1 &= -4 \pa_x \psi_3 \,\eps_2 \,\eps_4, &\qquad 
\pa_{\h}^2 \psi_3 &= -4 \pa_x \psi_1 \,\eps_2 \,\eps_4\, . \nonu
\end{alignat}
In terms of variables $u_{\pm}=u_1 \pm u_3$ and 
$\psi_{\pm}= \psi_1 \pm \psi_3$ this becomes 
\[\pa_{\h}^2 u_{\pm}  = \mp 4 \pa_x u_{\pm} \,\eps_2 \,\eps_4, 
\qquad \pa_{\h}^2 \psi_{\pm}  = \mp 4 \pa_x \psi_{\pm} \,\eps_2 \,
\eps_4 \, .\]
Each of elements in the kernel $\cK$ of $E$ in $\widehat{sl}_0 (2 \v 2)$
gives rise to a symmetry transformation of the supersymmetric 
$N=2$ mKdV hierarchy. We will now explore the structure of these
symmetry transformations. Recall, that all symmetry transformations
commute with isospectral deformations. 
The kernel elements among the algebra generators
\rf{redsl22} give rise to the bosonic symmetry transformations
$\d_{K_2^{(2n+1)}}$ and $\d_{K_1^{(2n+1)}}$
and the supersymmetry transformations of the form 
$\d_{ \eps F_1^{(2n+\frac{3}{2})} }$, 
$\d_{\eps  F_2^{(2n+\h)} } $,
$\d_{ \eps F_3^{(2n+\frac{3}{2})} }$ and
$\d_{\eps  F_4^{(2n+\h)} } $.
Based on algebraic relations of generators 
we find that the commutation relations between
the supersymmetry transformations  and (in general non-local)
bosonic symmetry transformations are:
\begin{alignat}{2}
\left[ \d_{K_i^{(2n+1)}}\, , \,\d_{\eps F_1^{(2 m +\frac{3}{2})}} \right]
&= (-1)^i \d_{\eps F_2^{(2(n +m + 1)+\frac{1}{2})}} ,
&\quad \left[ \d_{K_i^{(2n+1)}}\, , \,\d_{\eps F_3^{(2 m +\frac{3}{2})}} \right]
&= (-1)^{i+1} \d_{\eps F_4^{(2(n +m + 1)+\frac{1}{2})}} \nonu\\
\left[ \d_{K_i^{(2n+1)}}\, ,\, \d_{\eps F_2^{(2m +\frac{1}{2})}} 
\right] &= (-1)^i \d_{\eps F_1^{(2(n +m) + \frac{3}{2})}} 
, &\quad \left[ \d_{K_i^{(2n+1)}}\, ,\, \d_{\eps F_4^{(2m +\frac{1}{2})}} 
\right] &= (-1)^{i+1} \d_{\eps F_3^{(2(n +m) + \frac{3}{2})}} 
\nonu
\end{alignat}
for $i=1,2$.

The commutation relations of the  fermionic symmetry 
transformations takes a form of the following 
graded symmetry algebra :
\br
\left[ \d_{\eps F_k^{(2 n +\frac{3}{2})}} \, , \,
\d_{{\bar \eps} F_k^{(2 m +\frac{3}{2})}} \right] &=& 
 2 \d_{\eps {\bar \eps} E^{(2(n+m +1)+1}} , \;\;k=1,3 \nonu\\
\left[ \d_{\eps F_r^{(2 n +\frac{1}{2})}} \, , \,
\d_{{\bar \eps} F_r^{(2 m +\frac{1}{2})}} \right] &=& 
- 2 \d_{\eps {\bar \eps} E^{(2(n+m) +1)}} ,\;\;r=2,4 \nonu
\er
The above graded algebra represents the complete
non-local symmetry structure of the supersymmetric $N=2$ mKdV hierarchy.

We now turn our attention to the $N=2$ Sine-Gordon model.
The model is constructed in terms of the following
objects :
\br
A_0 &=& - {\pa_x} B \, B^{-1} = -{ \pa_x} \p_1 \, M_1^{(0)} -
{\pa_x} \p_3 M_3^{(0)} 
\lab{aazero}\\
B &=& e^{\p_1 M_1^{(0)} +\p_3 M_3^{(0)}} \lab{an2b}\\
A_{1/2} &=& \psi_1 G_1^{(1/2)} + \psi_3 G_3^{(1/2)} 
\lab{aahalf}\\
\jmath_{-1/2} &=& {\psi}_2 G_2^{(-1/2)} + {\psi}_4 G_4^{(-1/2)} 
\lab{ajhalf}\\
E^{(-1)} &=& K_1^{(-1)}+K_2^{(-1)} +I_2^{(-1)} \lab{aeminus}
\er
given in terms of generators from the 
$\widehat{ sl}_0 (2 \v 2)$ subalgebra \rf{redsl22}.
In addition, we use $D^{(\frac{1}{2})}$ as given in \rf{dshalf}.

We will use the following notation:
\[
{\bar \psi}_{\pm} =\psi_2 \pm \psi_4, \;\;\;
{\psi}_{\pm} =\psi_1 \pm \psi_3, \;\;\;
\p^{\pm} =\p_1 \pm \p_3 \, . 
\]
Making use of 
\be
\begin{split}
B j_{-1/2} B^{-1}&=e^{\p_1 M_1 +\p_3 M_3}\( \psi_2
G_2^{(-1/2)} + \psi_4 G_4^{(-1/2)}\)
e^{-\p_1 M_1 -\p_3 M_3} \\
&= G_2^{(-1/2)} \( \psi_2 \cosh \p_3 \cosh \p_1 -\psi_4 \sinh \p_3 \sinh \p_1\)
\\
&+ G_4^{(-1/2)} \( -\psi_2 \sinh \p_3\sinh \p_1+ \psi_4 \cosh \p_3 \cosh
\p_1 \)\\
&+F_1^{(-1/2)} \( -\psi_2 \sinh \p_1\cosh \p_3+ \psi_4 \cosh \p_1 \sinh \p_3 \)
\\
&+ F_3^{(-1/2)} \( \psi_2 \cosh \p_1 \sinh \p_3 -\psi_4 \sinh \p_1 \cosh \p_3\)
\end{split}
\lab{dmhalfc}
\ee
equations of motion \rf{pamjph} become 
\br
\pa_{-1} \psi_{\pm} &=& -2 {\bar \psi}_{\mp} \cosh \p_{\pm} \, .
\lab{n4p1chpsi}
\er
{}From equation \rf{paxjmh}
it follows that
\be
\pa_x {\bar \psi}_{\pm} = -2 \psi_{\mp} \cosh \p_{\pm}  
\lab{pabpsip}
\ee
In addition, we have
\br
\pa_{-1} \pa_x \p_{+}&=& 4  \sinh \p_+ \cosh \p_- 
- 4 \psi_+  {\bar \psi}_+ \sinh \p_{-}   
\lab{n4eq5}\\
\pa_{-1} \pa_x \p_{-}&= & 4     \cosh \p_+ \sinh \p_- 
- 4 \psi_-  {\bar \psi}_- \sinh \p_{+}  
\lab{n4eq6}
\er
obtained from equation \rf{pampaxbb}
using an identity :
\be
\begin{split}
B E^{(-1)} B^{-1}&=e^{\p_1 M_1 +\p_3 M_3}\( 
K_1^{(-1)}+K_2^{(-1)} +K_3^{(-1)}\)
e^{-\p_1 M_1 -\p_3 M_3} \\
&= K_1^{(-1)} \cosh (2 \p_1) +K_2^{(-1)} \cosh (2\p_3) 
\\
&- M_2^{(-1)}  \sinh (2\p_1) - M_4^{(-1)} \sinh (2 \p_3)
+K_3^{(-1)}\, .
\end{split}
\lab{dmmc}
\ee

The above equations of motion
are invariant under the supersymmetry transformations:
\be
\delta \p_{\pm} = 2 \psi_{\mp} \eps_{\pm},\qquad
\delta \psi_{\pm}= - \pa_x \p_{\mp} \eps_{\pm} 
\lab{sustrn2}
\ee
where
\[
\eps^{\pm} = \eps_2 \pm \eps_4 \,. 
\]
Furthermore from equation \rf{pahjmh}
we get
\be
\delta {\bar \psi}_{\pm}= 2 \sinh  \p_{\pm}  \eps_{\mp},
\lab{sustrn2c}
\ee
The above $N=2$ Sinh-Gordon equations  coincide with ones 
proposed by Kobayashi and 
Uematsu in \ct{Kobayashi:1991st,Kobayashi:1991jv} (see also the recent paper
\ct{Nepomechie:2001xk}) if we rescale $E$ and $\psi, \bar \psi $.

\section{\sf N=4 mKdV equation from subalgebra of $\widehat{sl}(4 \v 4)$  }
\label{section:n4mkdv}
Based on the  subalgebra of $\widehat{sl}(4,4)$  
generated by generators from formula (\ref{sub}) 
of Appendix \ref{section:sl44algebra} we define Lax operator
as in \rf{laxoper1}
where 
\[
A_0= u_1 M_1^{(0)}+ u_3 M_3^{(0)} +u_5 M_5^{(0)}+ u_7 M_7^{(0)} 
\]
\[
A_{\h} = {\bar \psi}_1 G_1^{(\h)} + {\bar \psi}_3 G_3^{(\h)}
+ {\bar \psi}_5 G_5^{(\h)} + {\bar \psi}_7 G_7^{(\h)}
\]
Notice that within the subalgebra generated by (\ref{sub})  
the kernel component $k_0$  in the Lax operator vanishes identically.
Next, we solve the zero-curvature equation \rf{zc1}
given for $n=3$  explicitly  by the commutation relation
\rf{zc3} with  $E^{(3)}= K_1^{(3)}+K_2^{(3)}+I^{(3)}$.

We found the following solution for the $D$'s :
\br
D^{(5/2)}&=& D^{(5/2)}_M = \l^2 A_{\h}\nonu \\
D^{(2)}_M&=& \l^2 A_{0} , \quad 
D^{(2)}_K = 0 \nonu\\
D^{(3/2)}_M &=& \h  \( \pa_x  \psi_1\, G_2^{(\frac{3}{2})} -
\pa_x  \psi_3\, G_4^{(\frac{3}{2})}-\pa_x  \psi_5\, G_6^{(\frac{3}{2})} +
\pa_x  \psi_7\, G_8^{(\frac{3}{2})}\)
\nonu \\
D^{(3/2)}_K &=& d_1 F_1^{(\frac{3}{2})}+ 
d_3 F_3^{(\frac{3}{2})}+ d_5 F_5^{(\frac{3}{2})}+ d_7 F_7^{(\frac{3}{2})}
\nonu \\
D^{(1)}_M&=&  c_2 M_2^{(1)}  + c_4 M_4^{(1)} +  c_6 M_6^{(1)}  + c_8 M_8^{(1)}  
\nonu \\
D^{(1)}_K&=& a_1 K_1^{(1)} +a_2 K_2^{(1)}+
a_3 K_3^{(1)} +a_6 K_6^{(1)} +a_7 K_7^{(1)}
+a_8 K_8^{(1)} +a_0 I^{(1)}
\nonu\\
D^{(1/2)}_M&=&  \b_1 G_1^{(\frac{1}{2})} +\b_3 G_3^{(\frac{1}{2})} +
\b_5 G_5^{(\frac{1}{2})} + \b_7 G_7^{(\frac{1}{2})} 
\nonu \\
D^{(1/2)}_K&=& \gamma_2 F_2^{(\frac{1}{2})}+\gamma_4 F_4^{(\frac{1}{2})}+
\gamma_6 F_6^{(\frac{1}{2})}+\gamma_8 F_8^{(\frac{1}{2})}
\nonu\\ 
D^{(0)}_M&=&  \d_1 M_1^{(0)} +\d_3 M_3^{(0)} +\d_5 M_5^{(0)} +\d_7 M_7^{(0)} 
\nonu \\
D^{(0)}_K&=&0
\er
where
\begin{alignat}{2}
d_1&=- \h \(\psi_1 u_1 +\psi_3 u_3 +\psi_5 u_5 +\psi_7 u_7 \)
, &\qquad
d_3&= \h \(\psi_1 u_3 +\psi_3 u_1 +\psi_5 u_7 +\psi_7 u_5 \)\nonu\\
d_5&= \h \(\psi_1 u_5 +\psi_3 u_7 +\psi_5 u_1 +\psi_7 u_3 \), &\qquad
d_7&=- \h \(\psi_1 u_7 +\psi_3 u_5 +\psi_5 u_3 +\psi_7 u_1 \)\nonu
\end{alignat}
\br
c_2&=& \h u_1^{\pr}-2\psi_1\(\psi_3u_3+\psi_5u_5\)+2 \psi_3\psi_7 u_5 
+2\psi_5\psi_7 u_3 \nonu\\
c_4&=& \h u_3^{\pr}+2\psi_1\(\psi_3u_1+\psi_5u_7\)-2 \psi_3\psi_7 u_7 
-2\psi_5\psi_7 u_1 \nonu\\
c_6&=& \h u_5^{\pr}+2\psi_1\(\psi_3u_7+\psi_5u_1\)-2 \psi_3\psi_7 u_1 
-2\psi_5\psi_7 u_7 \nonu\\
c_8&=& \h u_7^{\pr}-2\psi_1\(\psi_3u_5+\psi_5u_3\)+2 \psi_3\psi_7 u_3
+2\psi_5\psi_7 u_5 \nonu
\er
\br
a_0&=&-\psi_1 \pa_x \psi_1 - \psi_3 \pa_x \psi_3-\psi_5 \pa_x \psi_5
-\psi_7 \pa_x \psi_7\nonu\\
a_1&=&\psi_1 \pa_x \psi_1 - \psi_3 \pa_x \psi_3-\psi_5 \pa_x \psi_5
+\psi_7 \pa_x \psi_7-\h u_1^2-\h u_7^2 \nonu\\
a_2&=&-\psi_1 \pa_x \psi_1 + \psi_3 \pa_x \psi_3+\psi_5 \pa_x \psi_5
-\psi_7 \pa_x \psi_7-\h u_3^2-\h u_5^2 \nonu\\
a_3&=&\a_7= -\psi_1 \pa_x \psi_7 - \psi_3 \pa_x \psi_5-\psi_5 \pa_x \psi_3
-\psi_7 \pa_x \psi_1 \nonu\\
a_6&=& -\psi_1 \pa_x \psi_7 + \psi_3 \pa_x \psi_5+\psi_5 \pa_x \psi_3
-\psi_7 \pa_x \psi_1 -u_3u_5\nonu\\
a_8&=& \psi_1 \pa_x \psi_7 -\psi_3 \pa_x \psi_5-\psi_5 \pa_x \psi_3
+\psi_7 \pa_x \psi_1 -u_1u_7\nonu
\er
\br
\b_1&=& \frac{1}{4} \pa_x^2 \psi_1- 
\frac{1}{2} \psi_1  \( u_1^2+ u_3^2+ u_5^2+ u_7^2\)
-\h \psi_3  \(u_1u_3+u_5 u_7 \)\nonu \\
&-& \h \psi_5  \(u_1u_5+u_3 u_7 \) -\psi_7 \(u_3u_5+u_1 u_7 \)  
\nonu \\
\b_3&=& \frac{1}{4} \pa_x^2 \psi_3- 
\frac{1}{2} \psi_3  \( u_1^2+ u_3^2+ u_5^2+ u_7^2\)
-\h \psi_1  \(u_1u_3+u_5 u_7 \) \nonu \\
&-& \h \psi_7  \(u_1u_5+u_3 u_7 \) -\psi_5 \(u_3u_5+u_1 u_7 \)  
\nonu \\
\b_5&=& \frac{1}{4} \pa_x^2 \psi_5- 
\frac{1}{2} \psi_5  \( u_1^2+ u_3^2+ u_5^2+ u_7^2\)
-\h \psi_7  \(u_1u_3+u_5 u_7 \) \nonu \\
&-& \h \psi_1  \(u_1u_5+u_3 u_7 \) -\psi_3 \(u_3u_5+u_1 u_7 \)  
\nonu \\
\b_7&=& \frac{1}{4} \pa_x^2 \psi_7- 
\frac{1}{2} \psi_7  \( u_1^2+ u_3^2+ u_5^2+ u_7^2\)
-\h \psi_5  \(u_1u_3+u_5 u_7 \) \nonu \\
&-& \h \psi_3  \(u_1u_5+u_3 u_7 \) -\psi_1 \(u_3u_5+u_1 u_7 \)  
\nonu 
\er
\br
\gamma_2&=& \frac{1}{4} ( \psi_1 u_1^{\pr} - \psi_1^{\pr}  u_1
-\psi_3 u_3^{\pr} + \psi_3^{\pr}  u_3 - \psi_5 u_5^{\pr} + \psi_5^{\pr}  u_5
+\psi_7 u_7^{\pr} - \psi_7^{\pr}  u_7) \nonu\\
\gamma_4&=& \frac{1}{4} ( -\psi_1 u_3^{\pr} +\psi_1^{\pr}  u_3
+\psi_3 u_1^{\pr} - \psi_3^{\pr}  u_1 + \psi_5 u_7^{\pr} - \psi_5^{\pr}  u_7
-\psi_7 u_5^{\pr} + \psi_7^{\pr}  u_5) \nonu\\
\gamma_6&=& \frac{1}{4} ( -\psi_1 u_5^{\pr} + \psi_1^{\pr}  u_5
+\psi_3 u_7^{\pr} - \psi_3^{\pr}  u_7 +\psi_5 u_1^{\pr} - \psi_5^{\pr}  u_1
-\psi_7 u_3^{\pr} + \psi_7^{\pr}  u_3) \nonu\\
\gamma_8&=& \frac{1}{4} ( \psi_1 u_7^{\pr} - \psi_1^{\pr}  u_7
-\psi_3 u_5^{\pr} + \psi_3^{\pr}  u_5 - \psi_5 u_3^{\pr} + \psi_5^{\pr}  u_3
+\psi_7 u_1^{\pr} - \psi_7^{\pr}  u_1) \nonu
\er
\br
4\d_1&=&  \pa_x^2(u_1)-2 u_1 \(u_1^2+3u_7^2\) +3
u_1 \( \psi_1 \pa_x \psi_1 - \psi_3 \pa_x \psi_3-\psi_5 \pa_x \psi_5
+\psi_7 \pa_x \psi_7\) \nonu\\
&+&
3 u_3 \( -\psi_1 \pa_x \psi_3 + \psi_3 \pa_x \psi_1+\psi_5 \pa_x \psi_7
-\psi_7 \pa_x \psi_5\) \nonu\\ &+&
3 u_5 \(- \psi_1 \pa_x \psi_5 + \psi_3 \pa_x \psi_7+\psi_5 \pa_x \psi_1
-\psi_7 \pa_x \psi_3\) \nonu\\
&+& 3u_7 \(\psi_1 \pa_x \psi_7 - \psi_3 \pa_x \psi_5-\psi_5 \pa_x \psi_3
+\psi_7 \pa_x \psi_1\) 
\lab{delt1n4}\\
4\d_3&=& \pa_x^2(u_3)- 2 u_3 \(u_3^2+3u_5^2\) -
3u_3 \( \psi_1 \pa_x \psi_1 - \psi_3 \pa_x \psi_3-\psi_5 \pa_x \psi_5
+\psi_7 \pa_x \psi_7\) \nonu\\
&-&
3u_1 \( -\psi_1 \pa_x \psi_3 + \psi_3 \pa_x \psi_1+\psi_5 \pa_x \psi_7
-\psi_7 \pa_x \psi_5\) \nonu\\&-&
3 u_7 \(- \psi_1 \pa_x \psi_5 + \psi_3 \pa_x \psi_7+\psi_5 \pa_x \psi_1
-\psi_7 \pa_x \psi_3\) \nonu\\
&-&
3 u_5 \(\psi_1 \pa_x \psi_7 - \psi_3 \pa_x \psi_5-\psi_5 \pa_x \psi_3
+\psi_7 \pa_x \psi_1\) 
\lab{delt3n4}\\
4\d_5&=&  \pa_x^2(u_5)- 2u_5 \(u_5^2+3u_3^2\) -
3 u_5 \( \psi_1 \pa_x \psi_1 - \psi_3 \pa_x \psi_3-\psi_5 \pa_x \psi_5
+\psi_7 \pa_x \psi_7\) \nonu\\
&-&
3 u_7 \( -\psi_1 \pa_x \psi_3 + \psi_3 \pa_x \psi_1+\psi_5 \pa_x \psi_7
-\psi_7 \pa_x \psi_5\) \nonu\\&-&
3 u_1 \(- \psi_1 \pa_x \psi_5 + \psi_3 \pa_x \psi_7+\psi_5 \pa_x \psi_1
-\psi_7 \pa_x \psi_3\) \nonu\\
&-&
3 u_3 \(\psi_1 \pa_x \psi_7 - \psi_3 \pa_x \psi_5-\psi_5 \pa_x \psi_3
+\psi_7 \pa_x \psi_1\) 
\lab{delt5}\\
4\d_7&=&  \pa_x^2(u_7)- 2 u_7 \(u_7^2+3u_1^2\) +
3 u_7 \( \psi_1 \pa_x \psi_1 - \psi_3 \pa_x \psi_3-\psi_5 \pa_x \psi_5
+\psi_7 \pa_x \psi_7\) \nonu\\
&+&
3 u_5 \( -\psi_1 \pa_x \psi_3 + \psi_3 \pa_x \psi_1+\psi_5 \pa_x \psi_7
-\psi_7 \pa_x \psi_5\) \nonu \\&+&
3 u_3 \(- \psi_1 \pa_x \psi_5 + \psi_3 \pa_x \psi_7+\psi_5 \pa_x \psi_1
-\psi_7 \pa_x \psi_3\) \nonu\\
&+&
3 u_1 \(\psi_1 \pa_x \psi_7 - \psi_3 \pa_x \psi_5-\psi_5 \pa_x \psi_3
+\psi_7 \pa_x \psi_1\) 
\lab{delt7}
\er
This leads to the following equations of motion for fermions :
\br
4 \pa_3  \psi_1 &=& \pa_x^3 \psi_1- 
\frac{3}{2} \psi_1 \pa_x \( u_1^2+ u_3^2+ u_5^2+ u_7^2\)
- 3  \pa_x (\psi_1)  \( u_1^2+ u_3^2+ u_5^2+ u_7^2\)
\nonu\\
&-&3 \psi_3 \pa_x \(u_1u_3+u_5 u_7 \) -3 \psi_5 \pa_x \(u_1u_5+u_3 u_7 \) 
\nonu \\
&-&3 \psi_7 \pa_x \(u_3u_5+u_1 u_7 \) -6 \pa_x(\psi_7)  \(u_3u_5+u_1 u_7 \) 
\lab{xi1eqsmotn4}
\\
4 \pa_3  \psi_3 &=&
\pa_x^3 \psi_3- 
\frac{3}{2} \psi_3 \pa_x \( u_1^2+ u_3^2+ u_5^2+ u_7^2\)
- 3  \pa_x (\psi_3)  \( u_1^2+ u_3^2+ u_5^2+ u_7^2\)
\nonu\\
&-&3 \psi_1 \pa_x \(u_1u_3+u_5 u_7 \) -3 \psi_7 \pa_x \(u_1u_5+u_3 u_7 \) 
\nonu \\
&-&3 \psi_5 \pa_x \(u_3u_5+u_1 u_7 \) -6 \pa_x(\psi_5)  \(u_3u_5+u_1 u_7 \) 
\lab{xi3eqsmotn4}\\
4 \pa_3  \psi_5 &=&
\pa_x^3 \psi_5- 
\frac{3}{2} \psi_5 \pa_x \( u_1^2+ u_3^2+ u_5^2+ u_7^2\)
- 3  \pa_x (\psi_5)  \( u_1^2+ u_3^2+ u_5^2+ u_7^2\)
\nonu\\
&-&3 \psi_7 \pa_x \(u_1u_3+u_5 u_7 \) -3 \psi_1 \pa_x \(u_1u_5+u_3 u_7 \) 
\nonu \\
&-&3 \psi_3 \pa_x \(u_3u_5+u_1 u_7 \) -6 \pa_x(\psi_3)  \(u_3u_5+u_1 u_7 \) 
\lab{xi5eqsmot}\\
4 \pa_3  \psi_7 &=&
\pa_x^3 \psi_7- 
\frac{3}{2} \psi_7 \pa_x \( u_1^2+ u_3^2+ u_5^2+ u_7^2\)
- 3  \pa_x (\psi_7)  \( u_1^2+ u_3^2+ u_5^2+ u_7^2\)
\nonu\\
&-&3 \psi_5 \pa_x \(u_1u_3+u_5 u_7 \) -3 \psi_3 \pa_x \(u_1u_5+u_3 u_7 \) 
\nonu \\
&-&3 \psi_1 \pa_x \(u_3u_5+u_1 u_7 \) -6 \pa_x(\psi_1)  \(u_3u_5+u_1 u_7 \) 
\lab{xi7eqsmot}
\er
and bosonic equations of motion:
\be
\pa_3 u_i = \pa_x \d_i, \quad i=1,3,5,7
\lab{ueqsmotn4}
\ee
These are the $N=4$ supersymmetric mKdV equations.

They are invariant under the supersymmetry transformations derived from the 
zero-curvature equation \rf{zchalf}
with
\br
D^{(\frac{1}{2})} &= &\eps_2 F_2^{(1/2)} +\eps_4 F_4^{(1/2)}
+ \eps_6 F_6^{(1/2)} +\eps_8 F_8^{(1/2)} \lab{dshalfN4}\\
D^{(0)} &= &\bar a_1 M_1^{(0)} +\bar a_3 M_3^{(0)}+\bar a_5 M_5^{(0)}+\bar a_7 M_7^{(0)} 
\lab{dzmn4}
\er
with
\br
\bar a_1&=& 2 \(-\psi_1 \eps_2 +\psi_3 \eps_4 +\psi_5 \eps_6 -\psi_7 \eps_8 \)\nonu\\
\bar a_3&=& 2 \(-\psi_1 \eps_4 +\psi_3 \eps_2 +\psi_5 \eps_8 -\psi_7 \eps_6\)\nonu\\
\bar a_5&=& 2 \(-\psi_1 \eps_6 +\psi_3 \eps_8 +\psi_5 \eps_2 -\psi_7 \eps_4 
\)\nonu\\
\bar a_7&=& 2 \(-\psi_1 \eps_8 +\psi_3 \eps_6 +\psi_5 \eps_4 -\psi_7 \eps_2 \)\nonu
\er
The supersymmetry transformations are 
\[ \pa_{\h} u_i = \pa_x \bar a_i, \quad i =1,3,5,7
\]
and
\br
\pa_{\h} \psi_1&=& u_1 \eps_2 - u_3 \eps_4 -u_5 \eps_6 +u_7 \eps_8 \nonu\\
\pa_{\h} \psi_3&=& u_1 \eps_4 -u_3 \eps_2 -u_5 \eps_8 +u_7 \eps_6\nonu\\
\pa_{\h} \psi_5&=& u_1 \eps_6 -u_3 \eps_8 -u_5 \eps_2 +u_7 \eps_4 
\nonu\\
\pa_{\h} \psi_7&=& u_1 \eps_8 -u_3 \eps_6 -u_5 \eps_4 +u_7 \eps_2 
\nonu
\er

\section{\sf N=4 Sinh-Gordon Model from subalgebra of $\widehat{sl}(4 \v4)$  }
\label{section:n4sinhgordon}
Using the same subalgebra  (\ref{sub}) of $\widehat{ sl}(4 \v4)$ we find 
\br
A_0 &=& {\bar \pa} B \, B^{-1} = {\bar \pa} \p_1 \, M_1 + 
{\bar \pa} \p_3 M_3 +
{\bar \pa} \p_5 \, M_5 + {\bar \pa} \p_7 M_7 
\lab{aazeron4}\\
B &=& e^{\p_1 M_1 +\p_3 M_3+\p_5 M_5 +\p_7 M_7} \lab{an2bn4}\\
A_{1/2} &=& \psi_1 G_1^{(1/2)} + \psi_3 G_3^{(1/2)} +
\psi_5 G_5^{(1/2)} + \psi_7 G_7^{(1/2)} 
\lab{aahalfn4}\\
\jmath_{-1/2} &=& {\psi}_2 G_2^{(-1/2)} + {\psi}_4 G_4^{(-1/2)} 
+ {\psi}_6 G_6^{(-1/2)} + {\psi}_8 G_8^{(-1/2)} 
\lab{ajhalfn4}
\er
with $D^{(\frac{1}{2})}$ as in \rf{dshalfN4} and $E^{(-1)}$
as in \rf{aeminus}.

In notation:
\[
{\bar \psi}_{\pm} =\psi_2 \pm \psi_4, \;\;\;
{\psi}_{\pm} =\psi_1 \pm \psi_3, \;\;\;
\chi_{\pm} = \psi_5 \pm \psi_7\;\;\;
{\bar \chi}_{\pm} = \psi_6 \pm \psi_8
\]
\[
\p^{\pm} =\p_1 \pm \p_3, \;\;\;
{\bar \p}^{\pm} =\p_5 \pm \p_7
\]
equations of motion obtained from equation \rf{pamjph}
become : 
\br
\pa_{-1} \psi_{\pm} &=& -2 {\bar \psi}_{\mp} \cosh \p_{\pm} \cosh {\bar
\p}_{\pm} +2 {\bar \chi}_{\mp} \sinh \p_{\pm} \sinh {\bar \p}_{\pm}
\lab{n4p1chpsilc}\\
\pa_{-1} \chi_{\pm} &=& -2 {\bar \psi}_{\mp} 
\sinh \p_{\pm} \sinh {\bar \p}_{\pm} +2 {\bar \chi}_{\mp}\cosh \p_{\pm} 
\cosh {\bar \p}_{\pm}
\lab{n4p1chipm}
\er
{}From equation \rf{paxjmh}
it follows that
\br
\pa_x {\bar \psi}_{\pm} &=& -2 \psi_{\mp} \cosh \p_{\pm} \cosh {\bar \p}_{\pm}
+2 \chi_{\mp} \sinh \p_{\pm} \sinh {\bar \p}_{\pm} \lab{n4pabpsip}\\
\pa_x {\bar \chi}_{\pm} &=& 2 \chi_{\mp}  \cosh \p_{\pm} \cosh {\bar \p}_{\pm}
- 2 \psi_{\mp} \sinh \p_{\pm} \sinh {\bar \p}_{\pm} 
\lab{n4pabchip}
\er
{}From \rf{pampaxbb}
we obtain :
\be
\begin{split}
\pa_{-1} \pa_x \p_{+}&= 4 \left[ \sinh \p_+ \cosh \p_- 
\cosh {\bar \p}_{+} \cosh {\bar \p}_{-} -  \cosh \p_+ \sinh \p_- 
\sinh {\bar \p}_{+} \sinh {\bar \p}_{-}  \right]  \\
&+ 4 \psi_+ \( -{\bar \psi}_+ \sinh \p_{-} \cosh {\bar \p}_{-} +
{\bar \chi}_{+} \cosh \p_{-} \sinh {\bar \p}_- \) \\
&+ 4 \chi_+ \( {\bar \psi}_{+} \cosh \p_{-} \sinh {\bar \p}_-
-{\bar \chi}_+ \sinh \p_{-} \cosh {\bar \p}_{-}  \) 
\end{split}
\lab{n4eq5n4}
\ee
\be
\begin{split}
\pa_{-1} \pa_x \p_{-}&= - 4 \left[ \sinh \p_+ \cosh \p_- 
\sinh {\bar \p}_{+} \sinh {\bar \p}_{-} -  \cosh \p_+ \sinh \p_- 
\cosh {\bar \p}_{+} \cosh {\bar \p}_{-}  \right]  \\
&+ 4 \psi_- \( -{\bar \psi}_- \sinh \p_{+} \cosh {\bar \p}_{+} +
{\bar \chi}_{-} \cosh \p_{+} \sinh {\bar \p}_+ \) \\
&+ 4 \chi_- \({\bar \psi}_{-} \cosh \p_{+} \sinh {\bar \p}_+
-{\bar \chi}_- \sinh \p_{+} \cosh {\bar \p}_{+} \) 
\end{split}
\lab{n4eq6n4}
\ee
and
\be
\begin{split}
\pa_{-1} \pa_x {\bar \p}_{+}&= 4 \left[ \cosh \p_+ \cosh \p_- 
\sinh {\bar \p}_{+} \cosh {\bar \p}_{-} -  \sinh \p_+ \sinh \p_- 
\cosh {\bar \p}_{+} \sinh {\bar \p}_{-}  \right]  \\
&+ 4 \psi_+ \( {\bar \psi}_+ \cosh \p_{-} \sinh {\bar \p}_{-} -
{\bar \chi}_{+} \sinh \p_{-} \cosh {\bar \p}_- \) \\
&+ 4 \chi_+ \( -{\bar \psi}_{+} \sinh \p_{-} \cosh {\bar \p}_- 
+{\bar \chi}_+ \cosh \p_{-} \sinh {\bar \p}_{-} \) 
\end{split}
\lab{n4eq7}
\ee
\be
\begin{split}
\pa_{-1} \pa_x {\bar \p}_{-}&= 4 \left[ \cosh \p_+ \cosh \p_- 
\cosh {\bar \p}_{+} \sinh {\bar \p}_{-} -  \sinh \p_+ \sinh \p_- 
\sinh {\bar \p}_{+} \cosh {\bar \p}_{-}  \right]  \\
&+ 4 \psi_- \( {\bar \psi}_- \cosh \p_{+} \sinh {\bar \p}_{+} -
{\bar \chi}_{-} \sinh \p_{+} \cosh {\bar \p}_+ \) \\
&+ 4 \chi_- \( -{\bar \psi}_{-} \sinh \p_{+} \cosh {\bar \p}_+
+{\bar \chi}_- \cosh \p_{+} \sinh {\bar \p}_{+} 
 \) 
\end{split}
\lab{n4eq8}
\ee
These equations are invariant under supersymmetry transformations:
\[
\delta \p_{\pm} = 2 \psi_{\mp} \eps_{\pm} - 2\chi_{\mp} {\bar \eps}_{\pm},
\quad
\delta {\bar \p}_{\pm} = 2 \psi_{\mp} {\bar \eps}_{\pm} - 2\chi_{\mp} {\eps}_{\pm},
\]
and 
\[
\delta \psi_{\pm}= - \pa_x \p_{\mp} \eps_{\pm} + \pa_x {\bar \p}_{\mp}
{\bar \eps}_{\pm},
\quad
\delta \chi_{\pm}=  \pa_x {\bar  \p}_{\mp} \eps_{\pm} - \pa_x {\p}_{\mp}
{\bar \eps}_{\pm}
\]
where
\[
\eps^{\pm} = \eps_2 \pm \eps_4 , \quad 
{\bar \eps}_{\pm}= \eps_6 \pm \eps_8
\]
Furthermore from \rf{pahjmh} we get
\[\delta {\bar \psi}_{\pm}= 2 \sinh  \p_{\pm} \cosh {\bar \p}_{\pm} \eps_{\mp} 
-2 \cosh \p_{\pm} \sinh  {\bar \p}_{\pm} {\bar \eps}_{\mp},
\]
\[
\delta {\bar \chi}_{\pm}=  
2 \cosh  \p_{\pm} \sinh {\bar \p}_{\pm} \eps_{\mp} 
-2 \sinh \p_{\pm} \cosh  {\bar \p}_{\pm} {\bar \eps}_{\mp}
\]
In the limit 
\[ {\bar \p}_{\pm}=\chi_{\pm} = {\bar \chi}_{\pm}=0
\]
the above equations reproduce 
the equations of motion \rf{n4p1chpsi}, \rf{pabpsip},
\rf{n4eq5} and \rf{n4eq6} of $N=2$ model and corresponding 
supersymmetry
transformations \rf{sustrn2} and \rf{sustrn2c}.

\section{\sf Concluding Remarks}
\label{section:conlude}
We offer here some summarizing comments on results obtained in 
the previous sections.
Deriving extended supersymmetry structure without non-local
terms was accomplished here due to two technical but crucial
choices concerning the underlying algebraic structure.
One choice was to base our construction on the loop 
algebras $\widehat{sl} (n |m)$ with $n=m$ leading 
to the identity $I$ becoming a part of algebra.
The second choice concerned reduction process. It was made in such a way
as to ensure the Lax operator only contained  Cartan generators
among the zero-grade terms.
We now briefly point out consequences of these two steps.

First, notice  that the structure of the  supersymmetry 
transformation is 
directly related to $\cK_{\h}$; the $\h$ grade sector of the 
kernel $\cK$. 
For the $N=2$ case the relevant superalgebra is  $sl(2,2)$ 
which has 2 generators spanning $\cK_{\h}$, namely 
$F_2^{(\h)}$ and $ F_4^{(\h)}$  both squaring to $E$ modulo
the identity element $I$, i.e. 
\be
(F_2^{(\h)})^2 = -E, \quad \quad (F_4^{(\h)})^2 = E - 2I
\lab{ker}
\ee
For the $N=4$ we have determined that the relevant subalgebra is composed of a semi direct 
product of 
$U(1)$ with $sl(2,2)\otimes sl(2,2)$.  Each $sl(2,2)$ subalgebra 
generates a subsector of grade 
$\h$ of the kernel ${\cal K}_{\h}$ like \rf{ker} such that 
 ${\cal K}_{\h}$ is spanned by 
$F_2^{(\h)}, F_4^{(\h)}, F_6^{(\h)} $ and $ F_8^{(\h)}$ with 
\be
(F_2^{(\h)})^2 = (F_8^{(\h)})^2 =-E, \quad \quad (F_4^{(\h)})^2 = (F_6^{(\h)})^2 =E - 2I
\lab{ker1}
\ee
Since, $\d_{I}=0$ in the sense of the definition \rf{kirheq}
we see that all the algebra elements in equations \rf{ker}-\rf{ker1}
give rise to identical (up to the sign) symmetry transformations.

Moreover, we observe that the same pattern emerges for higher supersymmetries (e.g. $N=2n=8$) 
where  there is a natural decomposition 
of super Lie algebra $sl(2n, 2n)$  into semi direct product of
$U(1)$ with $\otimes ^n sl(2,2)$. 

Finally, going back to the reduction process we observe that 
within the relevant subalgebra structure 
the $k_0$ term in the Lax (2.14) vanishes identically. 
This is verified explicitly  for the $sl(2,2)$ case in equation \rf{laxoper1}
and also for the $sl(4,4)$ case. 
This is an
important feature, since  the non-zero 
$k_0$ leads to the undesired non-local supersymmetry transformations. 
This feature,   in fact, provided us with  a guideline for choosing the   
relevant subalgebra.

Presently, we are studying the conservation laws for these models according to refs. \ct{Aratyn:2000sm} and \ct{npskdv}.  
Moreover, the generalization to higher $N$ supersymmetric models, in particular for $N=8$, is under investigation.

\lskip
{\bf Acknowledgments}: HA acknowledges support from Fapesp and thanks the IFT-Unesp for its hospitality.
LHY, JFG and AHZ thank Capes and 
CNPq for  partial support.

\appendix
\section{\sf Appendix - The algebra of $\widehat{sl}(4,4)$}
\label{section:sl44algebra}
\subsection{\sf $\widehat{ sl}(4\v 4)$}
The roots of $sl(4 \v 4)$ are given by
\br 
\a_1 &=& e_1 - e_2, \quad \a_2=e_2 - e_3, \quad \a_3=e_3 - e_4, \quad \a_4=e_4 - f_1, \nonu \\
\a_5 &=& f_1 - f_2, \quad \a_6=f_2 - f_3, \quad \a_7=f_3 - f_4, \quad 
\er
where $e_i \cdot e_j = \d_{ij}, \quad  f_i \cdot f_j = -\d_{ij}$.  

The grade of step operators $E_{\b}^{(n)}$ of 
corresponding loop algebra is given by
\br
{\rm Grade} (E_{\b}^{(n)}) = 
n + \h (e_1-e_2+e_3-e_4+f_1-f_2+f_3-f_4)\cdot \b
\er
Define Kernel of $E$ to be:
\br
f_{1, \eta}^{(n+\h)} &=&  (\eta E_{\a_4}^{(n+\h)} + E_{-\a_4}^{(n+\h)}) + (\eta E_{\a_3+\a_4+\a_5}^{(n+\h)} + E_{-\a_3-\a_4-\a_5}^{(n+\h)}),
\nonu \\
f_{2, \eta}^{(n+\h)} &=&  (\eta E_{\a_3+\a_4}^{(n-\h)} + E_{-\a_3-\a_4}^{(n+\frac{3}{2})}) + 
(\eta E_{\a_4+\a_5}^{(n+\frac{3}{2})} + E_{-\a_4-\a_5}^{(n-\h)}),\nonu \\
f_{3, \eta}^{(n+\h)} &=&  (\eta E_{\a_2+\a_3+\a_4}^{(n+\h)} + E_{-\a_2-\a_3-\a_4}^{(n+\h)}) + 
(\eta E_{\a_1+\a_2+\a_3+\a_4+\a_5}^{(n+\h)} + E_{-\a_1-\a_2-\a_3-\a_4-\a_5}^{(n+\h)}),\nonu \\
f_{4, \eta}^{(n+\h)} &=&  (\eta E_{\a_4+\a_5+\a_6}^{(n+\h)} + E_{-\a_4-\a_5-\a_6}^{(n+\h)}) + 
(\eta E_{\a_3+\a_4+\a_5+\a_6+\a_7}^{(n+\h)} + E_{-\a_3-\a_4-\a_5-\a_6-\a_7}^{(n+\h)}),\nonu \\
f_{5, \eta}^{(n+\h)} &=&  (\eta E_{\a_1+\a_2+\a_3+\a_4}^{(n-\h)} + E_{-\a_1-\a_2-\a_3-\a_4}^{(n+\frac{3}{2})}) + 
(\eta E_{\a_2+\a_3+\a_4+\a_5}^{(n+\frac{3}{2})} + E_{-\a_2-\a_3-\a_4-\a_5}^{(n-\h)}),\nonu \\
f_{6, \eta}^{(n+\h)} &=&  (\eta E_{\a_4+\a_5+\a_6+\a_7}^{(n-\h)} + E_{-\a_4-\a_5-\a_6-\a_7}^{(n+\frac{3}{2})}) + 
(\eta E_{\a_3+\a_4+\a_5+\a_6}^{(n+\frac{3}{2})} + E_{-\a_3-\a_4-\a_5-\a_6}^{(n-\h)}),\nonu \\
f_{7, \eta}^{(n+\h)} &=&  (\eta E_{\a_2+\a_3+\a_4+\a_5+\a_6}^{(n+\h)} + E_{-\a_2-\a_3-\a_4-\a_5-\a_6}^{(n+\h)}) \nonu \\
&+& 
(\eta E_{\a_1+\a_2+\a_3+\a_4+\a_5+\a_6+\a_7}^{(n+\h)} + E_{-\a_1-\a_2-\a_3-\a_4-\a_5-\a_6-\a_7}^{(n+\h)}),\nonu \\
f_{8, \eta}^{(n+\h)} &=&  (\eta E_{\a_1+\a_2+\a_3+\a_4+\a_5+\a_6}^{(n-\h)} + E_{-\a_1-\a_2-\a_3-\a_4-\a_5-\a_6}^{(n+\frac{3}{2})})  \nonu \\
&+& (\eta E_{\a_2+\a_3+\a_4+\a_5+\a_6+\a_7}^{(n+\frac{3}{2})} + E_{-\a_2-\a_3-\a_4-\a_5-\a_6-\a_7}^{(n-\h)}),\nonu 
\er
for $\eta=\pm 1$.

\br
K_1^{(n)} &=& \( E_{\a_7}^{(n+1)}+E_{-\a_7}^{(n-1)} \) + \( E_{\a_5}^{(n+1)}+E_{-\a_5}^{(n-1)} \), \nonu \\
K_2^{(n)}&=& \( E_{\a_1}^{(n-1)}+E_{-\a_1}^{(n+1)} \) + \( E_{\a_3}^{(n-1)}+E_{-\a_3}^{(n+1)} \), \nonu \\
K_3^{(n)}&=&  -(\a_5 + 2\a_6 + \a_7)\cdot H^{(n)},  \nonu \\
K_6^{(n)}&=& \( E_{\a_1}^{(n-1)}+E_{-\a_1}^{(n+1)} \) - \( E_{\a_3}^{(n-1)}+E_{-\a_3}^{(n+1)} \), \nonu \\
K_7^{(n)}&=&(\a_1 + 2\a_2 +\a_3)\cdot H^{(n)}, \nonu \\
K_8^{(n)}&=& -\( E_{\a_7}^{(n+1)}+E_{-\a_7}^{(n-1)} \) + \( E_{\a_5}^{(n+1)}+E_{-\a_5}^{(n-1)} \), \nonu \\
I^{(n)} &=&  (\a_1 + 2\a_2 +3\a_3 + 4\a_4 + 3\a_5 + 2\a_6 + \a_7)\cdot H^{(n)},
 \nonu \\
K_9^{\pm})^{(n)} &=& (E_{\a_1+\a_2}^{(n)} \pm  E_{-\a_1-\a_2}^{(n)}) +  (E_{\a_2+\a_3}^{(n)} \pm  E_{-\a_2-\a_3}^{(n)}), \nonu \\
(K_{10}^{\pm})^{(n)} &=& (E_{\a_5+\a_6}^{(n)} \pm  E_{-\a_5-\a_6}^{(n)}) +  (E_{\a_6+\a_7}^{(n)} \pm  E_{-\a_6-\a_7}^{(n)}), \nonu \\
(K_{11}^{\pm})^{(n)} &=& (E_{\a_1+\a_2+\a_3}^{(n-1)} \pm  E_{-\a_1-\a_2-\a_3}^{(n+1)}) +  
(E_{\a_2}^{(n+1)} \pm  E_{-\a_2}^{(n-1)}), \nonu \\
(K_{12}^{\pm})^{(n)} &=& (E_{\a_5+\a_6+\a_7}^{(n+1)} \pm  E_{-\a_5-\a_6-\a_7}^{(n-1)}) +  
(E_{\a_5}^{(n+1)} \pm  E_{-\a_5}^{(n-1)}), 
\label{3.7} 
\er

and Image of $E$
\br
g_{1, \eta}^{(n+\h)} &=&  (\eta E_{\a_4}^{(n+\h)} + E_{-\a_4}^{(n+\h)}) - (\eta E_{\a_3+\a_4+\a_5}^{(n+\h)} + E_{-\a_3-\a_4-\a_5}^{(n+\h)}),
\nonu \\
g_{2, \eta}^{(n+\h)} &=&  (\eta E_{\a_3+\a_4}^{(n-\h)} + E_{-\a_3-\a_4}^{(n+\frac{3}{2})}) - 
(\eta E_{\a_4+\a_5}^{(n+\frac{3}{2})} + E_{-\a_4-\a_5}^{(n-\h)}),\nonu \\
g_{3, \eta}^{(n+\h)} &=&  (\eta E_{\a_2+\a_3+\a_4}^{(n+\h)} + E_{-\a_2-\a_3-\a_4}^{(n+\h)}) - 
(\eta E_{\a_1+\a_2+\a_3+\a_4+\a_5}^{(n+\h)} + E_{-\a_1-\a_2-\a_3-\a_4-\a_5}^{(n+\h)}),\nonu \\
g_{4, \eta}^{(n+\h)} &=&  (\eta E_{\a_4+\a_5+\a_6}^{(n+\h)} + E_{-\a_4-\a_5-\a_6}^{(n+\h)}) - 
(\eta E_{\a_3+\a_4+\a_5+\a_6+\a_7}^{(n+\h)} + E_{-\a_3-\a_4-\a_5-\a_6-\a_7}^{(n+\h)}),\nonu \\
g_{5, \eta}^{(n+\h)} &=&  (\eta E_{\a_1+\a_2+\a_3+\a_4}^{(n-\h)} + E_{-\a_1-\a_2-\a_3-\a_4}^{(n+\frac{3}{2})}) - 
(\eta E_{\a_2+\a_3+\a_4+\a_5}^{(n+\frac{3}{2})} + E_{-\a_2-\a_3-\a_4-\a_5}^{(n-\h)}),\nonu \\
g_{6, \eta}^{(n+\h)} &=&  (\eta E_{\a_4+\a_5+\a_6+\a_7}^{(n-\h)} + E_{-\a_4-\a_5-\a_6-\a_7}^{(n+\frac{3}{2})}) - 
(\eta E_{\a_3+\a_4+\a_5+\a_6}^{(n+\frac{3}{2})} + E_{-\a_3-\a_4-\a_5-\a_6}^{(n-\h)}),\nonu \\
g_{7, \eta}^{(n+\h)} &=&  (\eta E_{\a_2+\a_3+\a_4+\a_5+\a_6}^{(n+\h)} + E_{-\a_2-\a_3-\a_4-\a_5-\a_6}^{(n+\h)}) \nonu \\
&-& 
(\eta E_{\a_1+\a_2+\a_3+\a_4+\a_5+\a_6+\a_7}^{(n+\h)} + E_{-\a_1-\a_2-\a_3-\a_4-\a_5-\a_6-\a_7}^{(n+\h)}),\nonu \\
g_{8, \eta}^{(n+\h)} &=&  (\eta E_{\a_1+\a_2+\a_3+\a_4+\a_5+\a_6}^{(n-\h)} + E_{-\a_1-\a_2-\a_3-\a_4-\a_5-\a_6}^{(n+\frac{3}{2})})\nonu \\
& - &
(\eta E_{\a_2+\a_3+\a_4+\a_5+\a_6+\a_7}^{(n+\frac{3}{2})} + E_{-\a_2-\a_3-\a_4-\a_5-\a_6-\a_7}^{(n-\h)}),\nonu 
\er

\br
M_1^{(n)}&=& -(\a_5 + \a_7)\cdot H^{(n)}, \nonu \\
M_2^{(n)}&=& \( E_{\a_7}^{(n+1)}-E_{-\a_7}^{(n-1)} \) + \( E_{\a_5}^{(n+1)}-E_{-\a_5}^{(n-1)} \), \nonu \\
M_3^{(n)}&=& -(\a_1 + \a_3)\cdot H^{(n)}, \nonu \\
M_4^{(n)}&=& -\( E_{\a_1}^{(n-1)}-E_{-\a_1}^{(n+1)} \) - \( E_{\a_3}^{(n-1)}-E_{-\a_3}^{(n+1)} \), \nonu \\
M_5^{(n)}&=& (-\a_1 + \a_3)\cdot H^{(n)}, \nonu \\
M_6^{(n)}&=& -\( E_{\a_1}^{(n-1)}-E_{-\a_1}^{(n+1)} \) + \( E_{\a_3}^{(n-1)}-E_{-\a_3}^{(n+1)} \), \nonu \\
M_7^{(n)}&=& (-\a_5 + \a_7)\cdot H^{(n)}, \nonu \\
M_8^{(n)}&=& -\( E_{\a_7}^{(n+1)}-E_{-\a_7}^{(n-1)} \) + \( E_{\a_5}^{(n+1)}-E_{-\a_5}^{(n-1)} \), \nonu \\
(M_9^{\pm})^{(n)} &=& (E_{\a_1+\a_2}^{(n)} \pm  E_{-\a_1-\a_2}^{(n)}) -  (E_{\a_2+\a_3}^{(n)} \pm  E_{-\a_2-\a_3}^{(n)}), \nonu \\
(M_{10}^{\pm})^{(n)} &=& (E_{\a_5+\a_6}^{(n)} \pm  E_{-\a_5-\a_6}^{(n)}) -  (E_{\a_6+\a_7}^{(n)} \pm  E_{-\a_6-\a_7}^{(n)}), \nonu \\
(M_{11}^{\pm})^{(n)} &=& (E_{\a_1+\a_2+\a_3}^{(n-1)} \pm  E_{-\a_1-\a_2-\a_3}^{(n+1)}) -  
(E_{\a_2}^{(n+1)} \pm  E_{-\a_2}^{(n-1)}), \nonu \\
(M_{12}^{\pm})^{(n)} &=& (E_{\a_5+\a_6+\a_7}^{(n+1)} \pm  E_{-\a_5-\a_6-\a_7}^{(n-1)}) -  
(E_{\a_5}^{(n+1)} \pm  E_{-\a_5}^{(n-1)}), 
\label{3.6}
\er
\subsection{\sf Reduction of $\widehat{sl} (4 \v 4)$}
Reduction is performed in two steps. 

The first step consists of an algebraic reduction of $sl(4 \v 4)$ algebra
leading to a subalgebra of with Grassmaniann generators 
 written in the form (with $i=1, \ldots, 8 $):
\br
 R(\eta_i, \zeta_i)  &=& \( \eta_1 E_{\a_1+ \cdots + \a_4} + \zeta_1 E_{-\a_1- \cdots - \a_4}\) + 
\( \eta_2 E_{\a_1+ \cdots + \a_5} + \zeta_2E_{-\a_1- \cdots - \a_5}\) \nonu \\
&+& \( \eta_3 E_{\a_2+ \cdots + \a_4} + \zeta_3 E_{-\a_2- \cdots - \a_4}\) + 
\( \eta_4 E_{\a_2+ \cdots + \a_5} + \zeta_4 E_{-\a_2- \cdots - \a_5}\) \nonu \\
&+& \( \eta_5 E_{\a_3+ \cdots + \a_6} + \zeta_5 E_{-\a_3- \cdots - \a_6}\) + 
\( \eta_6 E_{\a_3+ \cdots + \a_7} + \zeta_6 E_{-\a_3- \cdots - \a_7}\) \nonu \\
&+& \( \eta_7 E_{\a_4+ \cdots + \a_6} + \zeta_7 E_{-\a_4- \cdots - \a_6}\) + 
\( \eta_8 E_{\a_4+ \cdots + \a_7} + \zeta_8 E_{-\a_4- \cdots - \a_7}\), 
\label{f}
\er
Denote 
\br
F_1 &=& R(\eta_i = 1, \zeta_i=1), \nonu \\
F_2 &=& R(\eta_i = -1, \zeta_i=1), \nonu \\
F_3 &=& R(\eta_1 = \eta_4=\eta_5=\eta_8 = \zeta_1 = \zeta_4=\zeta_5=\zeta_8 =-1, {\rm all \;\; other\;} \eta_i =\zeta_i=1), \nonu \\
F_4 &=& R(\eta_2 = \eta_3=\eta_6=\eta_7 = \zeta_1 = \zeta_4=\zeta_5=\zeta_8 =-1, {\rm all \;\; other\;} \eta_i =\zeta_i=1 ), \nonu \\
F_5 &=& R(\eta_1 = \eta_4=\eta_6=\eta_7 = \zeta_1 = \zeta_4=\zeta_6=\zeta_7 =-1, {\rm all \;\; other\;} \eta_i =\zeta_i=1 ), \nonu \\
F_6 &=& R(\eta_2 = \eta_3=\eta_5=\eta_8 = \zeta_1 = \zeta_4=\zeta_6=\zeta_7 =-1, {\rm all \;\; other\;} \eta_i =\zeta_i=1 ), \nonu \\
F_7 &=& R(\eta_5 = \eta_6=\eta_7=\eta_8 = \zeta_5 = \zeta_6=\zeta_7=\zeta_8 =-1, {\rm all \;\; other\;} \eta_i =\zeta_i=1 ), \nonu \\
F_8 &=& R(\eta_1 = \eta_2=\eta_3=\eta_4 = \zeta_5 = \zeta_6=\zeta_7=\zeta_8 =-1, {\rm all \;\; other\;} \eta_i =\zeta_i=1 ),
\label{a6}
\er
and 
\br
G_1 &=& R(\eta_2 = \eta_4=\eta_6=\eta_8 = \zeta_2 = \zeta_4=\zeta_6=\zeta_8 =-1, {\rm all \;\; other\;} \eta_i =\zeta_i=1), \nonu \\
G_2 &=& R(\eta_1 = \eta_3=\eta_5=\eta_7 = \zeta_2 = \zeta_4=\zeta_6=\zeta_8 =-1, {\rm all \;\; other\;} \eta_i =\zeta_i=1), \nonu \\
G_3 &=& R(\eta_1 = \eta_2=\eta_5=\eta_6 = \zeta_1 = \zeta_2=\zeta_5=\zeta_6 =-1, {\rm all \;\; other\;} \eta_i =\zeta_i=1), \nonu \\
G_4 &=& R(\eta_3 = \eta_4=\eta_7=\eta_8 = \zeta_1 = \zeta_2=\zeta_5=\zeta_6 =-1, {\rm all \;\; other\;} \eta_i =\zeta_i=1 ), \nonu \\
G_5 &=& R(\eta_1 = \eta_2=\eta_7=\eta_8 = \zeta_1 = \zeta_2=\zeta_7=\zeta_8 =-1, {\rm all \;\; other\;} \eta_i =\zeta_i=1 ), \nonu \\
G_6 &=& R(\eta_3 = \eta_4=\eta_5=\eta_6 = \zeta_1 = \zeta_2=\zeta_7=\zeta_8 =-1, {\rm all \;\; other\;} \eta_i =\zeta_i=1 ), \nonu \\
G_7 &=& R(\eta_2 = \eta_4=\eta_5=\eta_7 = \zeta_2 = \zeta_4=\zeta_5=\zeta_7 =-1, {\rm all \;\; other\;} \eta_i =\zeta_i=1 ), \nonu \\
G_8 &=& R(\eta_1 = \eta_3=\eta_6=\eta_8 = \zeta_2 = \zeta_4=\zeta_5=\zeta_7 =-1, {\rm all \;\; other\;} \eta_i =\zeta_i=1 ),
\label{a7}
\er
It is possible to show that the given generatos $F_i, G_i, M_i, \quad i=1, \cdots 8$, $K_a, K_{a+3}, a=1,2,3$ and $I$ 
of $sl(4 \v 4)$ closes into a subalgebra  using the following commutation relations
\br
\left[ H_i, H_j \right] &=& 0, \nonu \\
 \left[ H_i, E_a \right] &=& \a^i E_a, \nonu \\
\left[ E_{\a}, E_{\b} \right] &=& 
\left\{
\begin{matrix}
\eps (\a, \b)  E_{\a+\b},  & \a+\b = \; {\rm root}, \\
{{\a \cdot H}}, & \a+\b = \; {0}, \\
0,  &{\rm otherwise}
\end{matrix}
\right. \nonu
\er
where we 
$\eps (\a, \b)$ is the structure constant.

As the next second step 
we introduce a loop structure within such subalgebra by 
multiplying each generator 
by $\lambda^r$, where $r \in {\mathbb Z}$ for the
bosonic and $r \in {\mathbb Z} + \h$ fermionic generators,
respectively.
In the present paper we shall be using the following loop
reduction of the algebra defined by
\br
\{ M_1^{(r)}, M_3^{(r)}, M_5^{(r)}, M_7^{(r)}\}, \quad r = 2n, \nonu \\
\{M_2^{(r)}, M_4^{(r)}, M_6^{(r)}, M_8^{(r)},K_1^{(r)}, K_2^{(r)},
K_3^{(r)}, K_6^{(r)}, K_7^{(r)}, K_8^{(r)}, I^{(r)}\} 
\quad r = 2n+1, \nonu \\
\{G_1^{(r)}, G_3^{(r)}, G_5^{(r)}, G_7^{(r)}, F_2^{(r)}, F_4^{(r)}, 
F_6^{(r)}, F_8^{(r)}\} , \quad r = 2n + \h, \nonu \\
\{G_2^{(r)}, G_4^{(r)}, G_6^{(r)}, G_8^{(r)}, F_1^{(r)}, F_3^{(r)}, 
F_5^{(r)}, F_7^{(r)}\}, \quad r = 2n + \frac{3}{2},
\label{sub}
\er
where $n \in {\mathbb Z}$. 
The first four generators, namely $ M_1, M_3, M_5, M_7$ defined in 
equation (\ref{3.6}) involve the following 
 Cartan subalgebra elements
\br
\a_1\cdot H, \;\;  \a_3\cdot H, \;\;  \a_5\cdot H, \;\;  \a_7\cdot H
\label{2}
\er
while the kernel generators $K_3, K_7$ and $ I$ defined in 
equation (\ref{3.7}) involve the other three Cartan subalgebra generators 
\br
\a_2\cdot H, \; \; \a_4\cdot H, \;\;  \a_6\cdot H \,.
\label{3}
\er
The other generators $M_2, M_4, M_6, M_8, K_1,K_2, K_6, K_8$ 
give raise to the following step operators
\br
E_{\pm \a_1},\;  E_{\pm \a_3},\;  E_{\pm \a_5}, \; E_{\pm \a_7} \, .
\label{4}
\er
Finally, the 16 generators $G_i, F_i, i=1, \cdots 8$ in (\ref{a6}) 
and (\ref{a7}) are linear combinations of step operators :
\br
E_{\pm (\a_1 + \a_2 + \a_3 + \a_4)}, E_{\pm (\a_1 + \a_2 + \a_3 + \a_4+ \a_5)}, E_{\pm (\a_2 + \a_3 + \a_4)},
E_{\pm (\a_2 + \a_3 + \a_4+ \a_5)}, \nonu \\
E_{\pm ( \a_3 + \a_4+ \a_5 + \a_6)}, E_{\pm ( \a_3 + \a_4+ \a_5 + \a_6+ \a_7)},
E_{\pm ( \a_4+ \a_5 + \a_6)},E_{\pm (  \a_4+ \a_5 + \a_6+ \a_7)} \, .
\label{5}
\er
The key point we want to make here is that we may decompose the 
subalgebra in (\ref{sub}) into two commuting $sl(2,2)$ 
sectors shown below.

\subsection{$\b-sl(2,2)$}
\br 
 E_{\pm \b_1} &=& E_{\pm \a_1}, \nonu \\
E_{\pm \b_2} &=& E_{\pm (\a_2 + \a_3 + \a_4)},\nonu \\
E_{\pm \b_3} &=&  E_{\pm \a_5}, \nonu \\
E_{\pm (\b_1 +\b_2)} &=& E_{\pm (\a_1+\a_2 + \a_3 + \a_4)}, \nonu \\
E_{\pm (\b_2 +\b_3)} &=& E_{\pm (\a_2 + \a_3 + \a_4 +\a_5)}, \nonu \\
E_{\pm (\b_1+\b_2 +\b_3)}&=&  E_{\pm (\a_1+\a_2 + \a_3 + \a_4 +\a_5)}, \nonu \\
\b_1\cdot H &=& \a_1 \cdot H, \nonu \\
 \b_3\cdot H  &=& \a_5 \cdot H, \nonu \\
  \b_2 \cdot H &=& (\a_2 + \a_3 + \a_4)\cdot H 
\label{6}
\er
An identity element (in the sense that commutes with all 
generators in (\ref{6})) 
within such subalgebra is given by 
$I_{\b} = \(\a_1 + 2\a_2+2\a_3+2\a_4+\a_5\) \cdot H $.
The second $sl(2,2)$ sector is generated by 

\subsection{$\g-sl(2,2)$}
\br 
E_{\pm \g_1}  &=& E_{\pm \a_3}, \nonu \\
E_{\pm \g_2}  &=& E_{\pm (\a_4 + \a_5 + \a_6)},\nonu \\
E_{\pm \g_3}  &=& E_{\pm \a_7}, \nonu \\
E_{\pm (\g_1 +\g_2)} &=& E_{\pm (\a_1+\a_2 + \a_3 + \a_4)}, \nonu \\
E_{\pm (\g_2 +\g_3)} &=& E_{\pm (\a_4 + \a_5 + \a_6 +\a_7)}, \nonu \\
E_{\pm (\g_1+\g_2 +\g_3)}&=&  E_{\pm (\a_3+\a_4 + \a_5 + \a_6 +\a_7)}, \nonu \\
 \g_1 \cdot H&=& \a_3 \cdot H, \nonu \\
\g_3 \cdot H &=& \a_7 \cdot H, \nonu \\
 \g_2 \cdot H  &=& (\a_4 + \a_5 + \a_6)\cdot H 
\label{7}
\er
An identity element (in the sense that it commutes 
with all generators in (\ref{7})) 
within such subalgebra is given by 
$I_{\g} = \(\a_3 + 2\a_4+2\a_5+2\a_6+\a_7 \)\cdot H$.

Notice that the total identity (i.e. identity of $sl(4,4)$) is 
given as $I = I_{\b} + I_{\g}$, 
and therefore is not linearly independent of the
Cartan subalgebra generators in (\ref{6}) and (\ref{7}).  
However the subalgebra described by generators in 
(\ref{2}), (\ref{3}),(\ref{4}) and (\ref{5}) have 31 generators.  
Each $sl(2,2)$ subalgebra (\ref{6}) and (\ref{7}) have 
$15$ generators. 
There is one extra generator (which we may choose to be 
$\a_4\cdot H$ for symmetry) that does not commute 
with neither $sl(2,2)$ in (\ref{6}) or in  (\ref{7}).  
The relevant subalgebra consists therefore of a semi direct  
product of $U(1)$ with $sl(2,2)\otimes sl(2,2)$.

\end{document}